\documentclass[12pt]{emulateapj}
\slugcomment{{\em Accepted for publication in ApJ.}}

\usepackage{epsfig}
\usepackage{amsmath}
\usepackage{amssymb}
\usepackage{natbib}
\usepackage{graphicx}

\shorttitle{Equal- and unequal-mass mergers with black holes}
\shortauthors{Johansson et al.}

\begin{document}        

\title{Equal- and unequal-mass mergers of disk and elliptical galaxies with black holes}

\author{Peter H. Johansson\altaffilmark{1}, Thorsten Naab\altaffilmark{1}, Andreas Burkert\altaffilmark{1}}

\altaffiltext{1}{University Observatory Munich, Scheinerstr.\ 1, D-81679 Munich, Germany} 

\email{pjohan@usm.lmu.de}

\begin{abstract}

We present binary galaxy merger simulations with varying mass ratios 
and different progenitor morphologies. The simulations include mergers of gas-rich disks
(Sp-Sp), of early-type galaxies and disks (E-Sp, mixed mergers), and mergers of early-type galaxies 
(E-E, dry mergers).  We follow the dynamics of gas, stars and dark matter, and
include radiative cooling, star formation and black hole (BH) accretion and
the associated feedback processes as in Springel et al. We study  
the mass assembly of the BHs and discuss technical issues of the implemented model in detail. 
For Sp-Sp mergers, the peak star formation rate and BH accretion rate decrease
and the growth timescales of the central black holes and newly formed stars
increase with higher progenitor mass ratios. The peak BH accretion rate
typically occurs shortly after the time of BH merging for low progenitor mass
ratios (e.g. 3:1 and lower), whereas for higher progenitor mass ratios there is no clear
correlation between the peak BH accretion rate and BH merging time.
The termination of star formation
by BH feedback in disk mergers is significantly less important
for higher progenitor mass ratios (e.g. 3:1 and higher). In addition, the
inclusion of  BH feedback suppresses efficiently star formation in dry E-E mergers and mixed E-Sp mergers. 
All merger remnants, independent of their progenitors, follow the observed relations
between the central BH mass and the stellar velocity dispersion $(M_{{\rm
    BH}}-\sigma)$, the bulge mass $(M_{{\rm BH}}-M_{{*}})$ and the bulge
binding energy $(M_{{\rm BH}}-M_{*}\sigma^{2})$, with the dominant source of scatter
arising from variations in the initial gas mass fraction. The normalizations
for all relations and the simulated slope of the $M_{{\rm
    BH}}-\sigma$ and  $M_{{\rm BH}}-M_{*}\sigma^{2}$ relations are in good
agreement with the observations, whereas the simulated slope of the
$M_{{\rm BH}}-M_{{*}}$ relation is slightly steeper compared to the observations. This indicates that
the simple BH feedback model self-regulates the mass growth of the
BHs in accordance with the observed relations for a very wide range of merger progenitors.

\end{abstract}


\section{Introduction}

Observations in the last decade have shown that supermassive 
black holes (BH) reside in the centers of most if not all galaxies with spheroids 
(e.g. \citealt{1995ARA&A..33..581K}; \citealt{1998Natur.395A..14R}). 
The properties of these black holes and their host galaxies are correlated in
a number of ways, e.g. the correlation of the BH mass and the bulge velocity dispersion
as expressed in the $M_{\rm BH}-\sigma$ relation
(e.g. \citealp{2000ApJ...539L...9F}; \citealp{2000ApJ...539L..13G};
\citealp{2002ApJ...574..740T}), the relation between BH mass and the bulge stellar mass, $M_{{\rm BH}}-M_{\rm bulge}$
(e.g. \citealp{1998AJ....115.2285M}; \citealp{2003ApJ...589L..21M};
\citealp{2004ApJ...604L..89H}) and the relation between BH mass and the bulge
binding energy, $M_{{\rm BH}}-M_{\rm bulge} \sigma^{2}$
(e.g. \citealp{2007ApJ...665..120A}; \citealp{2007ApJ...662L..67B}). These observed relations indicate that 
supermassive BHs and the stellar spheroids of their host galaxies do not form
independently of each other.

There exists several theoretical explanations for the observed correlation
between galaxy properties and the black hole mass, including physical
mechanisms such as viscous disk accretion and star formation \citep{2001ApJ...554L.151B}, gas or
dark matter collapse \citep{2001ApJ...551L..31A}, dissipationless merging
\citep{2001ApJ...552L..13C} and the merging and migration of massive stellar
clumps and their associated intermediate-mass BHs to the galactic centers
(\citealp{2007ApJ...670..237B}; \citealp{2008arXiv0805.2266E}). The majority of the models rely on
some form of self-regulated BH mass growth, in which gas is fed to the central black
hole until the black hole releases sufficient energy to unbind the gas and
blow it away in momentum- or pressure-driven winds
(e.g. \citealp{1997ApJ...487L.105C,1998A&A...331L...1S,1999MNRAS.308L..39F,2005ApJ...618..569M,2007ApJ...665.1038C}). 

The first theoretical studies of the co-evolution of black holes and galaxies were
carried out using semi-analytic modeling (SAM) techniques
(e.q. \citealp{1999MNRAS.308...77C}; \citealp{2000MNRAS.311..576K};
\citealp{2003ApJ...595..614W}; \citealp{2003ApJ...582..559V};
\citealp{2004ApJ...600..580G}; \citealp{2006MNRAS.365...11C}; \citealp{2006MNRAS.370..645B}). In SAMs the
evolution of the dark matter halos is followed with merger trees
and the evolution of the galaxies and their embedded black holes is modeled
with simple physical parameterizations. Most SAMs are based on the assumption
that quasar activity is triggered by major galaxy mergers of gas-rich galaxies, in which large
quantities of gas flow to the centers of galaxies
(e.g. \citealp{1996ApJ...471..115B}; \citealp{2005ApJ...623L..67K};
\citealp{2006MNRAS.372..839N}; \citealp{2008ApJS..175..356H}) 
producing starbursts and leading to rapid black hole growth \citep{1989Natur.340..687H}.
During the mergers the dominant dynamical process is the violent relaxation
\citep{1967MNRAS.136..101L,2005MNRAS.362..252A} of the stellar and dark matter components.

This picture is supported by observations of local mergers that show large
central gas concentrations, accompanied by strong starburst and AGN
activity \citep{1998ApJ...498..579G}, indicating BH growth.
These so called ultraluminous infrared galaxies (ULIRGs) mostly originate from major mergers, typically with mass ratios below 3:1
(e.g. \citealp{2006ApJ...638..745D,2006ApJ...651..835D,2008MNRAS.tmp..110V}).
The ULIRGs could eventually evolve into spheroidal galaxies \citep{1999ApJ...522L..93H,2001ApJ...563..527G,2006AJ....132..976R,2006NewAR..50..720D}.
In addition, nearby quasars (e.g. \citealp{1999MNRAS.308..377M,2001ApJ...555..719C}) are preferentially
found in tidally disturbed objects, thus corroborating the important role of
galaxy interactions and mergers for major black hole growth. 
Furthermore, an increasing number of recent infrared and X-ray surveys have completed the
census of BH activity and mass growth by revealing a significant population of
highly obscured AGNs at redshifts above $z\sim 1$ 
\citep{2004A&A...427..795J,2005ApJ...622L.105H,2005AJ....129..578B}.

A first, simple self-consistent method for including black hole growth and thermal
feedback coupling to the surrounding gas in simulations of binary disk galaxy mergers was
developed by \citet{2005MNRAS.361..776S}. Based on this description 
hydrodynamical simulations of BH growth in galaxy mergers were used to
reproduce the local $M_{\rm BH}-\sigma$ relation \citep{2005Natur.433..604D}, study the redshift
evolution of this relation \citep{2006ApJ...641...90R}, provide an explanation
of the red colors of massive ellipticals \citep{2005ApJ...620L..79S}, and describe the
fundamental plane of elliptical galaxies \citep{2006ApJ...641...21R}. Using
this black hole merger sample \citet{2005ApJ...630..705H,2006ApJS..163....1H}
proposed a comprehensive picture of a unified merger-driven model for the
origin of starbursts, quasars and their relation to galaxy spheroid
formation. Recently, simulations of both individual galaxies \citep{2008MNRAS.387...13K}
and larger cosmological volumes \citep{2007MNRAS.380..877S,2008ApJ...676...33D} with BH feedback starting
directly from initial conditions appropriate for the $\Lambda$CDM cosmology
have been performed. Both cosmological studies assumed that seed black holes were
present at high redshifts in forming halos and then followed their evolution
self-consistently with hydrodynamical
simulations. \citet{2008ApJ...676...33D} employed the
\citet{2005MNRAS.361..776S} black hole feedback prescription, whereas
\citet{2007MNRAS.380..877S} used in addition a predominantly mechanical
feedback model for black holes in galaxy clusters with low accretion rates, in which AGN-driven bubbles
are injected into a gaseous environment \citep{2006MNRAS.366..397S}.
The study by \citet{2005Natur.433..604D} showed that the $M_{\rm BH}-\sigma$
relation can be reproduced for equal-mass disk galaxy mergers. The
cosmological studies of  \citet{2007MNRAS.380..877S} and \citet{2008ApJ...676...33D}
naturally include mergers with a range in mass ratios and morphologies, and both studies are
able to find adequate fits to the observed $M_{\rm BH}-\sigma$ and
$M_{{\rm BH}}-M_{{*}}$ relations at both high and low redshifts, where
$M_{{*}}$ is defined as the total stellar mass within the effective
radius. However, even the highest resolution cosmological simulations resolve
spatial scales that are an order of magnitude larger than the typical spatial scales
resolvable in high-resolution merger simulations $(\epsilon\sim 0.1 h^{-1}
\rm{kpc})$ and it is not obvious that low-resolution simulations that 
only resolve scales corresponding to the effective radii for typical
galaxies are able to capture the essential physics related to BH feedback.

In a cosmological context unequal-mass mergers are more frequent than
equal-mass mergers and the progenitors will have varying morphologies \citep{2001ApJ...561..517K,2006MNRAS.370..902K}.
It has been shown that the mass ratio of disk mergers has a significant
impact on the remnant properties. Equal-mass disk mergers predominantly result in
slowly rotating systems with boxy or disky isophotes, whereas unequal-mass
mergers lead to rotating, intrinsically anisotropic spheroids with disky isophotes
\citep{1998giis.conf..275B,1999ApJ...523L.133N,2001ApJ...554..291C,2002MNRAS.333..481B,2003ApJ...597..893N,2005A&A...437...69B,2005MNRAS.363..597B,2006ApJ...650..791C,2007arXiv0710.0663B}. 
The mass ratio and the actual gas content have a significant influence on the
final stellar orbit structure and internal kinematics of the merger remnants 
\citep{1996ApJ...471..115B,2005MNRAS.360.1185J,2007MNRAS.376..997J,2007MNRAS.381.1672T}.
Apparently, binary mergers of disk galaxies with varying mass ratios are one reasonable mechanism for the formation of
  intermediate- and low-mass ellipticals, whereas more massive slowly rotating
ellipticals are probably formed in multiple mergers at high redshifts
\citep{2007ApJ...658..710N,2007arXiv0710.0663B}. A potential assembly mechanism
for massive boxy and slowly-rotating elliptical galaxies at low redshift 
is merging of gas-poor early-type galaxies (dry mergers) \citep{2006ApJ...636L..81N}.
There is now mounting observational evidence for dry galaxy
mergers both in galaxy clusters \citep{1999ApJ...520L..95V,2005ApJ...627L..25T} and in the field 
\citep{2005AJ....130.2647V,2006ApJ...640..241B}. Simulations of
gas-free mergers have been shown to preserve the fundamental plane of
elliptical galaxies \citep{2003MNRAS.342..501N,2005MNRAS.362..184B}. 
Furthermore \citet{2006ApJ...641...21R} showed that E-E mergers
including BH feedback physics do not strongly alter the original $M_{\rm
  BH}-\sigma$ relation, although they might be the source for some of the
observed scatter in the relation. Mixed merging, in which a gas-poor early-type
  galaxy merges with a gas-rich spiral disk galaxy might also form elliptical
  galaxies with distinct properties. They were shown by
\citet{2003ApJ...597L.117K,2005MNRAS.359.1379K,2006ApJ...636L..81N,2008ApJS..175..390H} to dominate the merger budget
at intermediate masses between the low-mass ellipticals Sp-Sp mergers and the highest mass
E-E mergers. Thus it is of utmost importance to study whether the current BH
feedback model that reproduces the $M_{\rm BH}-\sigma$ and $M_{{\rm BH}}-M_{{*}}$ relations
for equal-mass disk mergers, is also able to reproduce the observed relations
in galaxy mergers with varying mass ratios and progenitor morphologies.

In this paper we present simulations of unequal-mass disk
mergers, mixed E-Sp mergers and E-E mergers, including BH feedback. Our aim is to study
the observed local $M_{\rm BH}-\sigma$ and $M_{{\rm BH}}-M_{{*}}$ relations
and see whether they hold for our merger sample with varying mass ratios and
different progenitor morphologies. 
We also study in detail
the effect of varying the merger mass ratio on the peak BH accretion rate and
final BH mass. In addition, we analyse the termination of star formation
by BH feedback in unequal-mass disk mergers, mixed E-Sp mergers and dry E-E
mergers compared to equal-mass disk mergers \citep{2005ApJ...620L..79S}. Furthermore, we discuss in detail the 
modeling of black hole mergers in the \citet{2005MNRAS.361..776S} feedback
prescription. The merging of black holes is not resolved in the numerical
simulations and therefore an effective model for BH merging needs to be adopted. 
We show that the assumption of rapid merging of the black holes in equal- and unequal-mass galaxy mergers is
critical for reproducing the final BH mass that fulfills the observed relations using the self-regulated
\citet{2005MNRAS.361..776S} BH feedback model.

This paper is structured as follows. In \S \ref{Methods} we describe the 
simulation code and discuss the BH feedback and merger prescriptions. 
The galaxy model and merger setups, together with the adopted parameter settings, are
discussed in \S \ref{Simulations}. Here we also simulate a set of
isolated disk galaxies and perform a detailed analysis of
the effect of the merger prescription and resolution on the growth and final
mass of the BHs. In \S \ref{Results} we present the results and
implications of our Sp-Sp, mixed E-Sp and dry E-E simulation sets to the origin of the $M_{\rm BH}-\sigma$,
the $M_{{\rm BH}}-M_{{*}}$ and the $M_{{\rm BH}}-M_{*}\sigma^{2}$
relations. In addition we perform here a detailed study of BH and star formation
activity as a function of merger mass ratio. Finally, we summarize and discuss
our findings in \S \ref{conc}.

\section{Methodology} 
\label{Methods}

\subsection{Numerical code}

We perform the simulations using the parallel TreeSPH-code GADGET-2
\citep{2005MNRAS.364.1105S}. The code follows the gas dynamics using the
Lagrangian Smoothed Particle Hydrodynamics (SPH)
(e.g. \citealp{1992ARA&A..30..543M}) technique formulated in such a way that 
energy and entropy is manifestly conserved \citep{2002MNRAS.333..649S}. 
Following \citet{1996ApJS..105...19K} the radiative cooling for a primordial
mixture of hydrogen and helium together with a spatially uniform
time-independent local UV background \citep{1996ApJ...461...20H} is included.

We include star formation and the associated supernova feedback, but exclude
supernova-driven galactic winds, following the sub-resolution multiphase
model developed by \citet{2003MNRAS.339..289S}. In this model the
ISM is treated as a two-phase medium in which cold clouds are embedded in
a tenous hot gas at pressure equilibrium. Stars form from the cold clouds in
regions were $\rho>\rho_{\rm th}$ with the shortlived stars supplying an
energy of $10^{51}$ ergs to the surrounding gas by supernovae. The threshold
density, $\rho_{\rm th}$, is determined self-consistently in the model by
requiring that the equation-of-state (EOS) is continuous at the onset of star
formation. 

We have implemented all the basic aspects of 
the \citet{2005MNRAS.361..776S}
effective BH feedback model into our version of GADGET-2. 
We summarize the main aspects of the BH feedback model 
in Appendix \ref{BH_feedback}. In Appendix
\ref{volker_comp} we perform a direct comparison of our feedback
implementation with the original model and find that the two models are in very
good agreement.

\subsection{Modeling black hole mergers}
\label{BH_merg}

In the final stages of a galaxy merger the dark matter halos of the galaxies
merge and in the center a single stellar system is formed from the coalescence
of the two stellar components. During this
process a substantial fraction of the gaseous component is funneled to the
center where it can drive strong starbursts and AGN activity. Presumably during
this process the two central supermassive BHs form a binary, that hardens and eventually 
leads to a coalescence of the BHs. However, it is not clear how long this
process would last and what the dominant mechanism for the hardening of the
binary would be. Possible mechanisms include hardening of the BH binary through
stellar-dynamical \citep{2004ApJ...602...93M,2005LRR.....8....8M} and
hydrodynamical \citep{2004ApJ...607..765E} processes. At the final stages of
the BH merger $(d\lesssim 0.01 \rm{pc})$ gravitational radiation will dominate angular
momentum and energy losses and cause the two BHs to coalesce
(e.g. \citealp{1998PhRvD..57.4535F}). However for unequal-mass BH pairs,
gravitational wave emission also removes net linear momentum and can impart a recoil
velocity to the center of mass of the system. This recoil velocity is
typically of the order of $v_{\rm CM}\sim 100-400 \ \rm{km s^{-1}}$ and may play
a role in removing relatively low-mass supermassive BHs from dwarf galaxies at
moderately to high redshifts \citep{2004ApJ...606L..17M,2004ApJ...607L...9M}.

In the simulations presented in this paper we lack the required resolution to
follow the hardening and eventual merging of the BHs. Using current numerical
techniques it is not possible to accurately resolve the 
dynamical evolution of a galactic BH pair on the relevant scales, although alternative techniques such as
the particle-splitting method \citep{2007Sci...316.1874M} and adaptive mesh
refinement simulations \citep{2008ApJ...678..154L} might allow one to follow
the BH dynamics to somewhat smaller separations compared to traditional methods.
Instead, as was the case for the BH
accretion, we use an effective model that ensures rapid merging of the BHs
once they come sufficiently close to each other. There is circumstantial evidence
suggesting that the merging of BHs is efficient. If this was not the case, it
should be common for a third black hole to be brought in by merger, in which
case the BH binary and third BH could lead to sling-shot ejection
of all three BHs \citep{1974ApJ...190..253S}. This process would seriously hamper the
growth of supermassive black holes. Thus the very existence of supermassive black
holes together with the relatively low number of known supermassive BH
binaries \citep{1988ApJ...325..628S,2005LRR.....8....8M} can be taken
as an indication that BH binaries typically merge efficiently. Upcoming
gravitational wave experiments, such as LISA will bring more clarity to the
duration and true number of BH mergers in the local Universe \citep{2002MNRAS.331..805H}.   

In this study we adopt the criterion of \citet{2005MNRAS.361..776S} for BH merging, which
allows for rapid BH merging. Two BHs are assumed
to merge instantly if they come within the smoothing length of each other and
if their relative velocity at this time is below the local sound speed, as
determined from the nearest neighbors within the SPH kernel. The momentum is
conserved when gas is absorbed by the BH sink particle, thus an
accreting BH moving in a gas-rich environment will experience a friction force
that reduces the relative motion of the BH with respect to the surrounding gas.
However, we
found that for unequal-mass mergers this description was
not always adequate in ensuring rapid merging of the black holes. Typically
the black holes had too large relative velocities to allow for merging and
thus an alternative method had to be employed to ensure the merging of the BHs.
After discussions with Volker Springel (Springel, private
communication) we decided to adopt a model, in which the BH is repositioned at 
every time step to the minimum of the local potential. This minimum is
evaluated by finding the gas particle within the SPH smoothing length of the
BH which has the minimum gravitational potential. The BH
is then repositioned to the location of this minimum at every timestep. 
This method effectively glues the BH to the center of the galaxy. Using this
description the BH will remain in the merging galaxies as long as the galaxy
core structure is intact. After disruption the two BHs will rapidly slide
to the center of the parent galaxy and merge. Although this method is crude it
should provide an adequate representation of a model, where one assumes that
BH merging is very efficient once the two BHs reside in the same parent galaxy.

\section{Simulations}
\label{Simulations}

\subsection{Galaxy models}
\label{galaxy_models}

Our progenitor galaxies are setup using the method described by \citet{2005MNRAS.361..776S}.
The virial mass and radius of each model is determined by the virial velocity
$v_{\rm vir}$ of the halo assuming a virial overdensity of $\Delta_{\rm vir}=200$ using 
the following relations

\begin{eqnarray}
M_{\rm vir}=\frac{v_{\rm vir}^{3}}{10 G H_{0}}, \\
r_{\rm vir}=\frac{v_{\rm vir}}{10 G H_{0}},
\end{eqnarray}
where $H_{0}=71 \ \rm{km s^{-1} Mpc^{-1}}$ is the present day Hubble
parameter. Thus $v_{\rm vir}$ sets the mass of the model. 
The virial radius together with the concentration parameter
$c=r_{\rm vir}/r_{\rm S}$ for a NFW halo \citep{1997ApJ...490..493N} is then
used to construct \citet{1990ApJ...356..359H} profile dark matter halos using the conversion detailed in 
\citet{2005MNRAS.361..776S}. For all galaxy models we assume a
concentration of $c=9$, which is a typical value for a halo at $z=0$ 
\citep{2001MNRAS.321..559B}. All of our model galaxies thus correspond to local $z=0$
disk galaxies and unlike \citet{2006ApJ...641...90R} we do not attempt a
redshift-dependent scaling of our initial conditions.

The dark matter halos are then populated with exponential disks with a
constant mass fraction of $m_{d}=0.041$ of the total
virial mass resulting in a total disk mass of $M_{d}=m_{d}M_{\rm vir}$ with a fractional gas content 
of $f_{\rm gas}$ and the rest being stars. The disk scale length $r_{d}$
is determined using the \citet{1998MNRAS.295..319M} formalism under the assumption that the
fractional disk angular momentum $j_{d}$ equals the disk mass fraction $m_{d}$ for a constant
halo spin of $\lambda=0.033$ for all models. This assumption
of $J_{d}=j_{d}J_{\rm halo}$ $(j_{d}=m_{d})$ corresponds to the conservation of the specific 
angular momentum of the material that forms the disk.
The value for the spin parameter
is close to the mean value found in numerical simulations
(e.g. \citealp{2002ApJ...581..799V}). The vertical scaleheight $z_{0}$ of the
stellar disk is radially constant and set to $0.2 r_{d}$ and the radial
velocity dispersion is set to equal the vertical velocity dispersion. The
vertical scale height of the gaseous disk is set via an integral constraint
on the surface mass density, a self-consistently determined galaxy
potential and the effective equation of state (EOS) of the multiphase
interstellar medium model \citep{1977ApJ...218..148M,2000MNRAS.317..697E,2006MNRAS.371.1519J}. 
For the EOS we adopt the strongly pressurized
multiphase model of the ISM, $q_{\rm EOS}=1$ \citep{2003MNRAS.339..289S} that
allows the construction of stable galaxy models even with very large gaseous fractions
\citep{2005MNRAS.361..776S}. 

All our galaxy models include a \citet{1990ApJ...356..359H} 
stellar bulge component with constant total mass fraction of $m_{b}=0.01367$ $(M_{b}=m_{b}M_{\rm vir})$, 
which is a third of the disk mass fraction. The bulge scale length $b$ is
also set to $0.2 r_{d}$. Finally we insert a seed black hole with mass
$M_{\rm BH}=10^{5} M_{\odot}$ at rest in the center of each galaxy model.

\subsection{Simulation parameters}
\label{sim_param}

At our standard numerical resolution we set the primary galaxy models up with 20,000 gas and
stellar disk particles, 10,000 bulge particles and 30,000 dark matter
particles. For the secondary galaxies in 3:1 mergers we use a third
of the particle resolution of the primary galaxies, thus resulting in 
6,667 gas and stellar disk particles, 3,333 bulge particles and 10,000 
dark matter particles. We set the gravitational softening length of gas, newly formed stars and the
black hole particles to $\epsilon=0.1 h^{-1} \rm{kpc}$ and scale the softening
lengths of the disk, bulge and more massive dark matter particles with the
square root of the particle mass resulting in  $\epsilon=0.2 h^{-1} \rm{kpc}$
for the bulge and disk particles and $\epsilon=0.8 h^{-1} \rm{kpc}$ for the
dark matter particles. This ensures that the maximum gravitational
force exerted from a particle is independent of its mass \citep{2001MNRAS.324..273D}. 
\citet{2006MNRAS.372..839N} also showed that small variations in the
respective softening lengths have only a minor effect on 
the central gas content, with typically only a 10\% increase in the central gas
content when equal softening lengths and mass resolution were employed for all
collisionless particles. The SPH properties of the gas and BH particle are averaged over the usual
GADGET-2 spline SPH
kernel using $\sim64$ SPH gas particles. Furthermore we require that the
minimum SPH smoothing length is equal to the gravitational softening
length. This choice safeguards against artificial stabilization of small clumps
at low resolution \citep{1997MNRAS.288.1060B}.   
In order to resolve the flow of gas around the black hole accurately we ran all
simulations with high force accuracy of $\alpha_{\rm force}=0.005$ and time
integration accuracy of $\eta_{\rm acc}=0.02$ (see \citealp{2005MNRAS.364.1105S} for details). 

The parameters governing the multi-phase feedback model are set as follows: 
The star formation timescale $t_{*}^{0}=8.4 \ \rm Gyr$, the cloud evaporation
parameter $A_{0}=4000$ and the supernova 'temperature' $T_{\rm{SN}}=4.0 \times
10^{8} \ \rm{K}$ that reflects the heating rate from a population of supernovae
for a given IMF. Thus, we here again follow \citet{2005MNRAS.361..776S} who
showed that these choices result in a star formation rate of $\sim 1
M_{\odot} \rm{yr^{-1}}$ for a Milky Way-like galaxy. However, we note that
these choices yield a comparatively low star formation rate that lies slightly
below the observed Kennicutt relation \citep{1998ARA&A..36..189K}.

We set the dimensionless efficiency parameter of accretion in 
Eq. \ref{Bondi} to $\alpha=100$, which is considerably higher than the
theoretically expected value of $\alpha\sim 1$. This discrepancy is
essentially due to limits imposed by the numerical resolution and
the effective subresolution model for star formation. The estimated values for
the gas density and thermal sound speed are large-scale averages of the gas at
the location of the BH resulting in artifically low densities and high sound
speeds compared to the case where the multi-phase structure of the ISM could
be spatially resolved on the scale of the Bondi radius. Thus the larger value
for $\alpha$ can essentially be seen as an empirical correction factor that
translates from the resolvable low mean density to the time-averaged small-scale
density at the location of the BH. A similar conclusion was reached by
\citet{2008MNRAS.387...13K}, who set $\alpha=300$ in their simulations.
The exact choice of $\alpha$ within a factor of a
few is not critical as long as $\alpha$ is large enough to allow a black hole
in a gas-rich environment to reach the Eddington regime, (again see
\citealp{2005MNRAS.361..776S} for details). As detailed in \S \ref{BH_feedback} 
we set the radiative efficiency to $\epsilon_{\rm{r}}=0.1$ and the thermal
coupling constant of the black hole feedback energy to $\epsilon_{f}=0.05$.

All simulations presented in this paper were evolved for a total of $t=3 \ \rm
Gyr$ using the local Altix 3700 Bx2 machine hosted at the University Observatory in Munich.

\subsection{Isolated disks with BHs}
\label{isolated_BH}

\begin{figure}
\centering 
\epsscale{0.75}
\includegraphics[width=8cm]{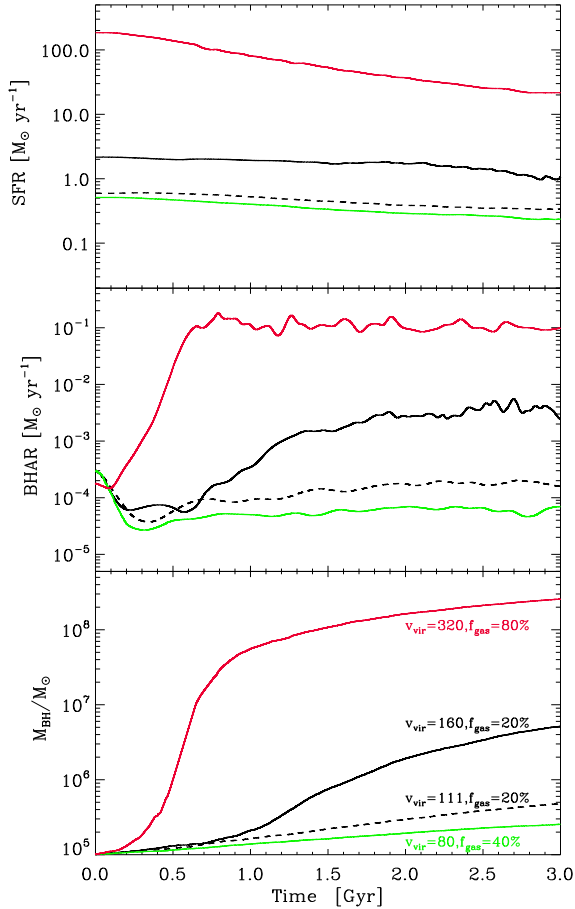}
\caption{The star formation rate (top), the black hole accretion rate (middle)
and the evolution of the black hole mass (bottom) as a function of time for
four isolated galaxy models with initial gas mass fractions of 20\% (black),
40\% (green) and 80\% (red) and virial velocities as indicated in the Figure.}
\label{SFR_BHacc_iso}
\end{figure}

\begin{table}
\caption{Isolated galaxies and the $M_{\rm BH}-\sigma$ and $M_{{\rm BH}}-M_{{*}}$ relations}             
\label{iso_sims}      
\centering          
\begin{tabular}{c c c c c c c c c c}
\hline\hline       
                      
$v_{\rm vir}$ \footnote{Virial velocity in $\rm{[kms^{-1}]}$} & $f_{\rm gas}$ & $M_{\rm BH}$\footnote{Final mass of the BH
  in $10^{5} M_{\odot}$} & $\sigma_{\rm bulge}$\footnote{Final bulge stellar
  velocity dispersion in km/s} &  $M_{\rm bulge}$\footnote{Final bulge mass in $10^{10} M_{\odot}$} & $\frac{M_{\rm{BH}}}{M_{\rm{BH},\sigma}}$\footnote{The ratio of the
  final BH mass to the BH mass expected from Eq. \ref{eq:MBH-sigma} if the
  final BH would lie on the $M_{\rm BH}-\sigma$ relation.} & $\frac{M_{\rm{BH}}}{M_{\rm{BH},M_{*}}}$\footnote{The ratio of the
  final BH mass to the BH mass expected from Eq. \ref{eq:MBH-M*} if the
  final BH would lie on the $M_{\rm BH}-M_*$ relation} \\ 
\hline                    
80  & 0.4 & 5.12 & 57.0 & 0.23  & 0.59 & 0.22 \\ 
111 & 0.2 & 6.50 & 78.3 & 0.61  & 0.21 & 0.09 \\
160 & 0.2 & 56.0 & 120.4 & 1.83 & 0.32 & 0.24 \\
320 & 0.8 & 2640 & 252.9 & 14.7 & 0.76 & 1.08 \\
\hline                  
\end{tabular}

\end{table}

We begin by running a representative sample of our disk galaxy models in isolation in order to
study the stability of the constructed galaxy models. The models are run at
our standard numerical resolution. In
Fig. \ref{SFR_BHacc_iso} we show the resulting star formation rates, BH accretion
rates and BH mass growth for four models with 20\% (black), 40\% (green), 80\%
(red) initial gas mass fractions. All galaxies show stable evolution
in their star formation and BH accretion rates, reproducing the isolated disk results 
for a stiff EOS with $q_{\rm EOS}=1$ presented by \citet{2005MNRAS.361..776S}. 

All models start with a seed black hole with mass
$M_{\rm BH}=10^{5} M_{\odot}$ at rest in the center, which grows due to gas
accretion during the simulation as is shown in the bottom panel of
Fig. \ref{SFR_BHacc_iso}. After completing the isolated disk simulations we
calculate the final black hole mass of the galaxies together with the
mass-weighted line-of-sight stellar velocity dispersion $\sigma_{\rm bulge}$ measured
from the bulge stellar particles using 50 randomly
projected realizations of the galaxy models.
Using the extracted velocity
dispersions we then calculate the expected BH masses of the galaxies and check if they
lie on the observed $M_{\rm BH}-\sigma$ relation,

\begin{equation}
\log (M_{\rm BH}/M_{\odot})=a+b \log (\sigma/\sigma_{0}),
\label{eq:MBH-sigma}
\end{equation}
where the relation is defined relative to $\sigma_{0}=200 \ \rm{km
s^{-1}}$. The observed values as determined by \citet{2002ApJ...574..740T} are
$a=8.13\pm0.06$ for the normalization coefficient and $b=4.02\pm0.32$ for the slope.
We also evaluate the total bulge stellar mass $M_{\rm bulge}$ by adding up the
mass of all the stellar bulge particles. By definition this mass is not
changing throughout the simulation, as the newly formed stars form from gas in
the disk and we do not include their potential contribution to the final
stellar bulge mass. 
We then compare the final BH masses with the
expected BH masses if the galaxies would lie on the observed 
$M_{{\rm BH}}-M_{*}$ relation, where we now use $M_{*}$ to denote the total
bulge mass.

\begin{equation}
\log (M_{\rm BH}/M_{\odot})=c+d \log (M_*/10^{11} M_{\odot}).
\label{eq:MBH-M*}
\end{equation}
The best fit observed values as determined by \citet{2004ApJ...604L..89H} are 
$c=8.20\pm0.10$ for the normalization coefficient and $d=1.12\pm0.06$ for the slope.
The final BH masses, stellar velocity dispersions, final stellar masses and the ratios between the
final BH mass and the BH mass derived from Eqs. \ref{eq:MBH-sigma} and
\ref{eq:MBH-M*} assuming that the galaxies lie on the $M_{\rm BH}-\sigma$ and $M_{{\rm BH}}-M_{*}$
relations respectively are tabulated in Table \ref{iso_sims}. 

The initial seed mass of $M_{\rm BH}=10^{5} M_{\odot}$ corresponds to a
velocity dispersion of $\sigma=33 \ \rm{km s^{-1}}$ (Eq. \ref{eq:MBH-sigma}), which would be the
corresponding velocity dispersion of a galaxy with $v_{\rm vir}\sim 50 \ \rm{km s^{-1}}$.
Thus the BHs of the galaxies are not initially on the $M_{\rm BH}-\sigma$
relation, but evolve towards this relation during the simulation. The bulge velocity
dispersions of the galaxies increase only slightly on the $\sim10\%$ level
during the simulations. The governing factor determining the evolution of the BHs for the
isolated disk galaxies is the depth of the potential well and the gas mass
fraction. This can be seen in the ratio of the final BH mass to the expected BH mass
from Eq. \ref{eq:MBH-sigma} tabulated in Table \ref{iso_sims}.
The lower the gas mass fraction the lower
the final BH masses are with respect to the expected $M_{\rm BH}-\sigma$ relation. Models
with lower gas mass fractions would require additional gas inflow to the central
regions in order to ensure sufficient BH growth. Such additional gas inflow 
could be triggered by galaxy mergers. The same initial BH seed mass of $M_{\rm
  BH}=10^{5} M_{\odot}$ would require initially a total bulge mass of $M_{*}=1.4\cdot 10^{8} M_{\odot}$. 
Again this would only be valid for the very lowest mass initial galaxy models
and thus generally the galaxies do not initially lie on the $M_{{\rm BH}}-M_{*}$ relation.
Typically the simulations of the isolated galaxies result in BH
masses that are too low for their given stellar bulge masses. Again an additional
central gas flow, potentially induced by a galaxy merger is required for the
central BHs to reach the masses expected by their corresponding stellar bulge masses. 
 
Interestingly the simulation of the most massive and gas-rich galaxy (the $f_{\rm gas}=80\%$, $v_{\rm vir}\sim 320 \ \rm{km s^{-1}}$ model
in Fig. \ref{SFR_BHacc_iso} and Table \ref{iso_sims}) results in a final
BH mass that is relatively close to the BH mass expected from the $M_{\rm
  BH}-\sigma$ relation and even superceeds the BH mass expected from the $M_{{\rm BH}}-M_{*}$ relation.
However, some of this BH mass growth
may be artificially high, due to the fact that the current BH feedback model
does not restrict the size of the BH accretion radius. Thus the accretion radius
of a a massive BH in a gas-rich region may grow to unphysically large sizes
and accrete material that would physically not end up in the BH because of its
high angular momentum and large distance from the BH. This is typically not a problem in 
merger simulations, where most of the BH mass grows by accretion of nearby
high-density gas.

\subsection{The effect of resolution on the $M_{\rm BH}-\sigma$ relation}
\label{merg_pre_mbh_sig}

\begin{figure}
\centering 
\includegraphics[width=9cm]{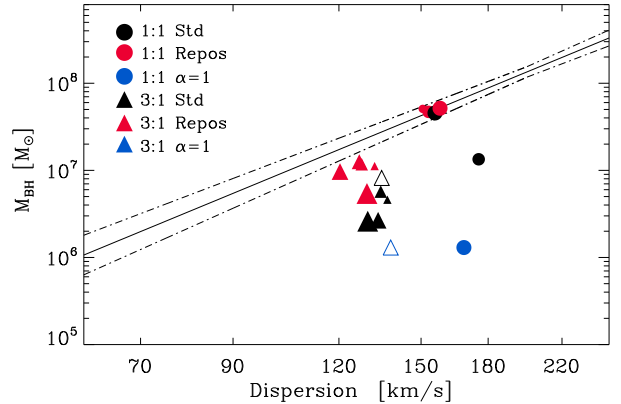}
\caption{The $M_{\rm BH}-\sigma$ relation as a function of resolution and
  merger prescription, with larger/smaller symbols indicating higher/lower
  resolution, respectively (Table \ref{31_res}). Triangles show 3:1 and circles 1:1 merger remnants. Black
  symbols employ the standard merger prescription and the red symbols use
  the BH repositioning method. The blue symbols also employ the BH
  repositioning method but with a reduced accretion efficiency of $\alpha=1$.
  The solid/open symbols indicate cases where the
  BHs did/did not merge. (Table \ref{31_res}). The lines
  show the observed relation with errors by \citet{2002ApJ...574..740T}.}
\label{MBH-mergpre}
\end{figure}

\begin{table*}
\caption{3:1 and 1:1 binary BH merger resolution and merger prescription study}             
\centering   
\scriptsize{
\label{31_res}      
\begin{tabular}{c c c c c c c c c c c c}
\hline\hline       
                      
Resolution\footnote{The secondary has a third of the particles of the primary}
  & Mass ratio & $\epsilon_{\rm{gas}}$\footnote{Gravitational softening
  lengths in $h^{-1}
  \rm{kpc}$}\footnote{$\epsilon_{\rm{gas}}=\epsilon_{\rm{stars}}=\epsilon_{\rm{BH}}$}
  & $\epsilon_{\rm{disk}}$\footnote{$\epsilon_{\rm{disk}}=\epsilon_{\rm{bulge}}$} & $\epsilon_{\rm{DM}}$
  & $N_{\rm{gas}}$\footnote{$N_{\rm{gas}}=N_{\rm{disk}}$} & $N_{\rm{bulge}}$  & $N_{\rm{DM}}$ 
  & $m_{\rm{gas}}$\footnote{Mass of a gas particle in $10^{5} M_{\odot}$} & 
 $M_{\rm BH,fin}$\footnote{Final mass of the BH in $10^{5} M_{\odot}$} & 
 $\sigma_{\rm fin}$\footnote{Final stellar velocity dispersion within $r_{\rm
  eff}$ in km/s} & 
 $t_{\rm merg}$\footnote{Time of BH merger in Gyr} \\

\hline                    
Tiny Std  & 3:1 & 0.16 & 0.32 & 1.28 & 5000 & 2500 & 7500 & $31$ &  46.0 & 136.9 & 2.02  \\ 
Tiny Repos &  3:1 & 0.16 & 0.32 & 1.28 & 5000 & 2500 & 7500 & $31$ & 118 & 132.2 & 2.81   \\
Low Std  & 3:1 & 0.125 & 0.25 & 1.00 & 10000 & 5000 & 15000 &  $15.5$ & 57.0 & 134.4 & 2.00 \\  
Low Repos  & 3:1 & 0.125 & 0.25 & 1.00 & 10000 & 5000 & 15000 & $15.5$ & 118 & 127.7 & 2.50  \\  
Standard Std  & 3:1 & 0.1 & 0.2 & 0.8 & 20000 & 10000 & 30000 & $7.75$ & 81.7 & 134.7 & ---  \\
Standard Repos & 3:1 & 0.1 & 0.2 & 0.8 & 20000 & 10000 & 30000 & $7.75$ & 126 & 126.8 & 1.82   \\
Medium Std  & 3:1 & 0.08 & 0.16 & 0.64 & 40000 & 20000 & 60000 & $3.87$ & 26.7 & 133.5 & 1.84   \\
Medium Repos  & 3:1 & 0.08 & 0.16 & 0.64 & 40000 & 20000 & 60000 & $3.87$ & 96.9 & 120.3 & 1.72  \\
High Std  & 3:1 & 0.0625 & 0.125 & 0.5 & 80000 & 40000 & 120000 &  $1.94$ & 26.1  & 129.8 & 2.81  \\
High Repos & 3:1 & 0.0625 & 0.125 & 0.5 & 80000 & 40000 & 120000 & $1.94$ & 54.2  & 129.5 & 1.86  \\
Tiny Std  & 1:1 & 0.16 & 0.32 & 1.28  & 5000 & 2500 & 7500 & $31$ & 420 & 155.4 &  2.21 \\
Tiny Repos &  1:1 & 0.16 & 0.32 & 1.28 & 5000 & 2500 & 7500 & $31$ & 508 & 150.7 & 1.57 \\
Low Std  & 1:1 & 0.125 & 0.25 & 1.00 & 10000 & 5000 & 15000 &  $15.5$ & 134 & 175.4 & 1.84  \\
Low Repos  & 1:1 & 0.125 & 0.25 & 1.00 & 10000 & 5000 & 15000 & $15.5$ & 469 &  153.1 & 1.57  \\
Standard Std  & 1:1 & 0.1 & 0.2 & 0.8 & 20000 & 10000 & 30000 & $7.75$ & 456 &  155.8 &  1.72  \\
Standard Repos & 1:1 & 0.1 & 0.2 & 0.8 & 20000 & 10000 & 30000 & $7.75$ & 516 & 157.9 &  1.50 \\
\hline                  
\end{tabular} 
}
\end{table*}

We then proceed with a detailed analysis of the effect of the BH merger
prescription and resolution on the final mass of BHs and the corresponding stellar velocity
dispersions in 3:1 and 1:1 merger remnants. We simulate 3:1 
mergers with a quarter (tiny), half (low), double (medium)
and four times (high) the number of particles compared to our standard
numerical resolution (\S \ref{sim_param}). In addition we compute a
control sample of 1:1 mergers at the standard, low and tiny resolutions. The
five 3:1 and three 1:1 initial conditions are simulated using both 
the standard BH merging prescription and the BH repositioning method described 
in \S \ref{BH_merg} resulting in a total of 16 simulations (see
Table \ref{31_res} for details). The gravitational softening lengths for 
all the particles including the BH are scaled with the inverse cube root of the number
of particles, with larger numerical resolution resulting in smaller
gravitational softening lengths. In addition we run a 3:1 and a 1:1 merger
with a lower accretion efficiency of $\alpha=1$ at our standard resolution
using the BH repositioning method.

All galaxy models have a 20\% initial gas fraction, with $v_{\rm vir,p}=160
\ \rm{kms^{-1}}$, $v_{\rm vir,s}=111 \ \rm{kms^{-1}}$ and are setup with the
parameters described in \S \ref{galaxy_models} (p and s denote the primary 
and secondary galaxies, respectively). In order to isolate the
effects of resolution and merging prescription all 3:1 and 1:1 mergers are run with
identical merging geometries and initial disk orientations. 
For the orbit and initial orientation we adopt geometry G13 for both the 3:1 mergers 
and 1:1 mergers from \citet{2003ApJ...597..893N}. The G13 geometry
corresponds to the inclinations $i_{p}=-109, i_{s}=180$ and the arguments of the
pericenter $\omega_{p}=60, \omega_{s}=0$ for the primary and secondary
galaxies, respectively. The galaxies approach each other on a parabolic orbit where the initial separation
of the progenitors is $R_{\rm init}=0.5(r_{\rm vir,p}+r_{\rm vir,s})$ 
and the pericentric distance is $r_{\rm peri}=r_{\rm d,p}+r_{\rm d,s}$,
where $r_{\rm vir,p}$, $r_{\rm d,p}$ and $r_{\rm vir,s}$, $r_{\rm d,s}$ are
the virial and disk scale radii for the primary and secondary galaxies, respectively. 
We note that radial parabolic or near-parabolic orbits are generally motivated by
statistics from N-body simulations \citep{2006A&A...445..403K}.

After completing the merger simulations we calculate the final black hole mass
together with mass-weighted line-of-sight stellar velocity dispersion $\sigma$
measured from all stellar particles within the projected half-mass radius
$r_{\rm e}$ using 50 randomly projected
realizations of the merger remnant. This determination of $\sigma$ is similar
to the method used by observers who
measure the luminosity-weighted velocity dispersion within an effective radius
\citep{2000ApJ...539L..13G}. The simulated final BH masses and stellar
velocity dispersions together with the time of the BH merger are tabulated in 
Table \ref{31_res}. 

For the 3:1 mergers, the merging prescription has a strong effect on the final
BH mass of the merger remnant, with the standard prescription producing
BH masses that are a factor of 2-3 lower than the corresponding
BH masses using the repositioning method. The final stellar velocity
dispersion on the other hand converges with increasing resolution, with
the repositioning method resulting in marginally lower velocity dispersions.
The 3:1 mergers with the standard BH merger prescription result in BH masses
that are too low for their corresponding velocity dispersions as shown in 
Fig. \ref{MBH-mergpre}, where the simulated 3:1 (triangles) and 1:1 (circles) results
are overplotted on the observed $M_{\rm BH}-\sigma$ relation from Eq. \ref{eq:MBH-sigma}
by \citet{2002ApJ...574..740T}.  

For the 1:1 mergers the effect of the merger prescription is less pronounced.
With the exception of a single simulation at low resolution,
all 1:1 mergers result in very similar BH masses and velocity dispersions. 
At the standard resolution the two prescriptions 
produce final BH masses and velocity dispersions that agree within 10\% with
each other and the observed relation.
All the 1:1 mergers with repositioning and two out of three of the standard 
prescription mergers lie on the observed $M_{\rm BH}-\sigma$ relation as shown
in Fig. \ref{MBH-mergpre}. 

Finally we show that reducing the accretion efficiency from our standard value
of $\alpha=100$ to the physically motivated $\alpha=1$ produces BH masses that are far too low for
their corresponding velocity dispersions (the blue points in
Fig. \ref{MBH-mergpre}). However, this is expected given the limitations
imposed by the numerical resolution and subresolution modeling as was discussed
in \S \ref{sim_param}. Changing the accretion efficiency or even removing
the BH altogether does not affect the final stellar velocity dispersion, which is
primarily set by the depth of the potential well. Thus, this implies that the
BH is not dynamically important for the galaxy as a whole. A similar problem would
arise when comparing the simulated BH masses to the total stellar
masses of the galaxies. Variations in numerical resolution and BH merging prescription affect only
marginally the star formation rate and thus the BH masses in the standard
prescription 3:1 merger remnants would be too low compared to their corresponding stellar masses.

The numerical resolution tests show that the central
stellar velocity dispersion converge within 10\% with increasing resolution,
whereas the BH mass converges within 25\%, with the exception of the highest
resolution simulation using the repositioning method. The highest resolution
repositioning run produces a BH mass, which is lower by a factor of two compared
to the other simulations. This is mainly due to the reduction of the
gravitational softening of the BH particle. The accretion rate and mass growth
of the BH is determined by a combination of the accretion efficiency ($\alpha$
in Eq. \ref{Bondi}) and the gravitational softening. Our chosen value of
accretion efficiency was adopted for our standard numerical resolution, thus
it is not surprising that smaller gravitational softening lengths produce
slightly smaller BH masses.

In Appendix \ref{merg_pre_res} we study in more detail the differences between the standard
and repositioning merger prescription for both 3:1 and 1:1 binary galaxy
mergers with BHs. We find that the repositioning merger prescription at the 
standard numerical resolution produces stable results and thus we adopt this prescription and numerical resolution 
for all remaining simulations presented in this paper.

\begin{figure}
\centering 
\includegraphics[width=9cm]{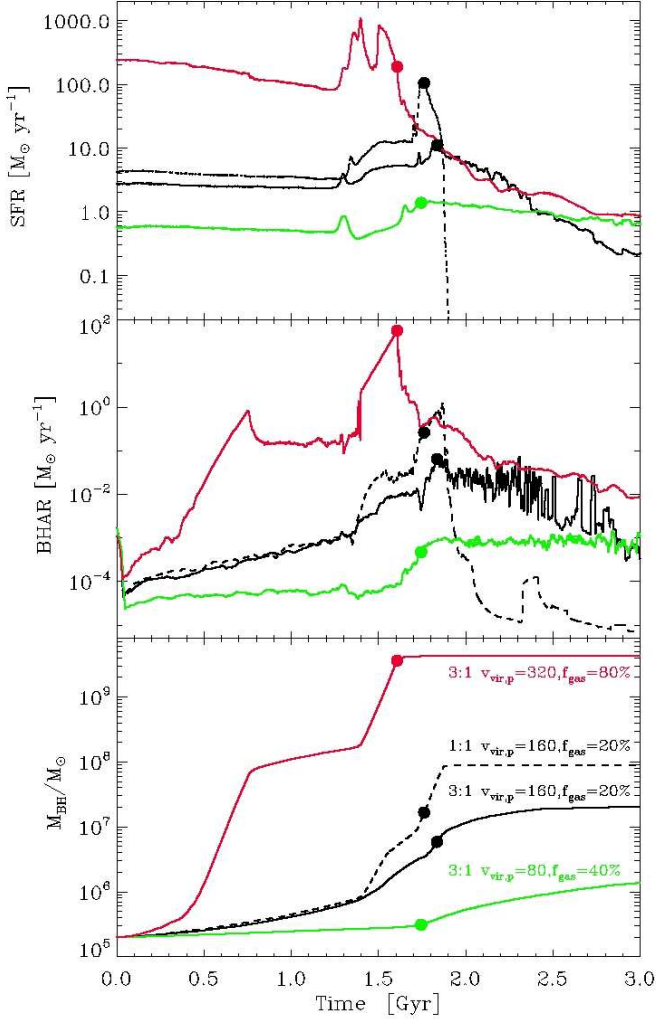}
\caption{The total star formation rate (top), the total black hole accretion rate (middle)
and the evolution of the total black hole mass (bottom) as a function of time for
three 3:1 (solid lines) and one 1:1 (dashed line) merger with initial gas mass 
fractions of 20\% (black), 40\% (green) and 80\% (red). The filled circles
indicate the time of merging of the BHs.} 
\label{SFR_BHacc_31}
\end{figure}

\begin{figure}
\centering 
\includegraphics[width=9cm]{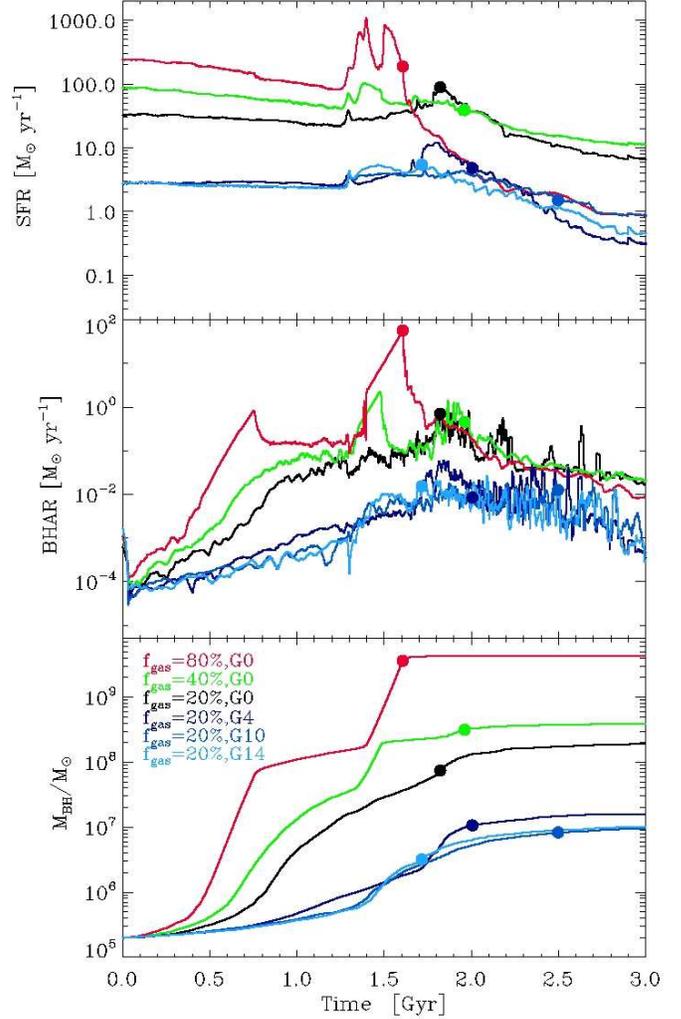}
\caption{The total star formation rate (top), the total black hole accretion rate (middle)
and the evolution of the total black hole mass (bottom) as a function of time for
three 3:1 mergers with varying initial gas mass fractions (black, green, red
lines) and three 3:1 mergers with varying orbital and initial geometries for a
fixed gas fraction (blue lines).}
\label{SFR_BHacc_geo_gas_31}
\end{figure}

\section{Results}
\label{Results}

\subsection{Unequal-mass 3:1 disk mergers with BHs}
\label{uneq_eq}

\begin{table*}
\caption{Unequal- and equal-mass mergers}             
\centering 
\tiny{
\label{sims}      
\begin{tabular}{c c c c c c c c c c c}
\hline\hline       
                      
Models &  $v_{\rm vir,prog1}$\footnote{All virial velocities in $\rm{kms^{-1}}$}
&  $v_{\rm vir,prog2}$ & Mass ratio & Orbit & $f_{\rm gas}$ 
& $N_{\rm{gas,tot}}$\footnote{$N_{\rm{gas}}=N_{\rm{disk}}$}  & $N_{\rm{bulge,tot}}$  &
  $N_{\rm{DM,tot}}$ & $N_{\rm{sim}}$ \\
\hline                    
S1-1:1  & 50, 80, 160, 320, 500 &  50, 80, 160, 320, 500 & 1:1 & G0 & 0.2, 0.4, 0.8 & 40000 & 20000 & 60000  
 & 15 \\  
S2-3:1 & 50, 80, 160, 320, 500 & 35, 55, 111, 222, 347 & 3:1 & G0 & 0.2, 0.4, 0.8 &  26667 & 13333  & 40000 
  & 15 \\  
G1-1:1  & 160 & 160 & 1:1 & G7, G10, G13 & 0.2 & 40000 & 20000 & 60000  
  & 3 \\  
G2-3:1 & 160 & 111 & 3:1 & G4, G10, G14 & 0.2 &  26667 & 13333  & 40000 
  & 3 \\  

\hline                  
\end{tabular}
}
\end{table*}

\begin{table}
\caption{Best fit $M_{\rm BH}-\sigma$ relation for 3:1 and 1:1 mergers}             

\label{mbh_sigma_31}      
\centering          
\begin{tabular}{c c c c c}
\hline\hline       
                      
Sample & N & $a$ & $b$ & $\Delta_{\log M_{\rm BH}}$ \\
\hline                    
Tot sample & 36 & $8.07\pm0.06$ & $3.82\pm0.15$ & 0.29 \\
3:1 sample & 18 & $8.06\pm0.08$ & $3.78\pm0.18$ & 0.33 \\
1:1 sample & 18 & $8.05\pm0.07$ & $3.77\pm0.18$ & 0.26 \\
S1-S2 20\% gas sample & 10 & $7.85\pm0.04$ & $3.47\pm0.12$ & 0.13 \\
S1-S2 40\% gas sample & 10 & $8.13\pm0.05$ & $3.96\pm0.13$ & 0.14 \\
S1-S2 80\% gas sample & 10 & $8.35\pm0.10$ & $3.77\pm0.28$ & 0.29 \\
Observed sample\footnote{From \citet{2002ApJ...574..740T}.} & 31 & $8.13\pm0.06$ & $4.02\pm0.32$ & 0.25-0.3 \\

\hline                  
\end{tabular}
\end{table}

\begin{figure*}
\centering 
\includegraphics[width=18cm]{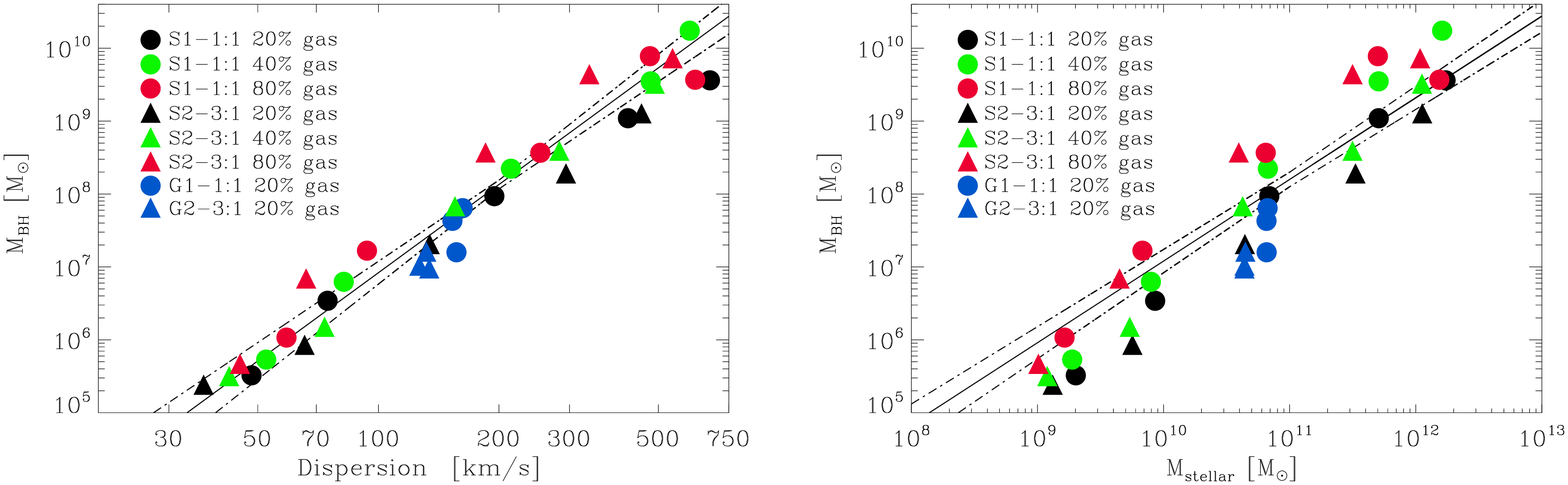}
\caption{The $M_{\rm BH}-\sigma$  (left) and $M_{{\rm BH}}-M_{*}$ (right) relations for our complete sample of 3:1
  (triangles) and 1:1 (circles) mergers. The black, green and red symbols show
  the effect of varying the initial gas mass fraction for our S1- (equal-mass) and
  S2-samples (unequal-mass). The blue symbols
  show the variation caused by the orbit and initial geometry for our G1- and G2-samples
  (see also Table \ref{sims}). The lines
  show the observed relations with errors by \citet{2002ApJ...574..740T} (left
  panel) and \citet{2004ApJ...604L..89H} (right panel), respectively.}
\label{MBH-sigma_31}
\end{figure*}

Using the galaxy models described in \S \ref{galaxy_models} we set up
five unequal-mass 3:1 mergers and five equal-mass 1:1 mergers with
three different initial gas mass fractions of $f_{\rm gas}=0.2, 0.4, 0.8$,
resulting in a total of thirty simulations (the S1- and S2-series in Table \ref{sims}). The progenitor galaxies are at the
standard numerical resolution with the virial velocities of the models ranging
from $v_{\rm vir,p}=50-500 \ \rm{km s^{-1}}$ and $v_{\rm vir,s}=35-347 \
\rm{km s^{-1}}$ for the primary and secondary galaxies, respectively. 
The progenitors merge on prograde, nearly parabolic, in plane orbits (G0
orbit), with inclinations $i_{1}=i_{2}=0$ and arguments of the pericenter 
$\omega_{1}=\omega_{2}=0$. 
The initial separation and
pericentric distances are set to the mean of the virial radius and twice the
mean of the scale radius respectively, as detailed in \S \ref{merg_pre_mbh_sig}.
In addition we simulate a set of three 3:1 and three 1:1 mergers with varying initial disk geometries
and orbits. These simulations (the G1- and G2-series in Table \ref{sims}) were all run
with $f_{\rm gas}=0.2$ and $v_{\rm vir,p}=160 \ \rm{km s^{-1}}$, $v_{\rm vir,s}=111 \
\rm{km s^{-1}}$ progenitors on the G7, G10, G13 orbits for the equal-mass mergers
and the G4, G10, G14 orbits for the unequal-mass mergers (see
\citealp{2003ApJ...597..893N} for a definition of the orbital parameters). The
orbits in the G-series were chosen to provide a good coverage of the expected 
anisotropy-ellipticity $(\delta-\epsilon)$ plane for merger remnants (see
\citealp{2007arXiv0710.0663B} for definitions of $\delta$ and $\epsilon$).

In Fig. \ref{SFR_BHacc_31} we plot the total star formation rates, BH accretion rates and
the BH mass growth for three 3:1 mergers (solid lines) with  20\% (black), 40\% (green), 80\%
(red) initial gas mass fractions together with one 20\% gas fraction 1:1 merger (dashed lines).
The star formation is very efficiently terminated by the BH feedback in 1:1 mergers, compared
to a generally much shallower decline in star formation for 3:1 mergers. In the
lowest mass 3:1 merger the BH feedback is actually unable to terminate the
star formation and the galaxy is forming stars at a constant rate even after
the BH merger. A similar trend is true for the BH accretion rates, with a
much gentler decline in the accretion rate for galaxies undergoing 3:1
mergers compared to 1:1 mergers. This evolution is mirrored in the growth of
the BH masses, with the final BH mass being lower in 3:1 mergers typically by
a factor of 2-5, but with a relatively large scatter depending on the
progenitor masses and initial gas fractions. Most of the mass growth of the
BHs typically takes place shortly after the merging of the BHs as indicated by the
filled circles in Fig. \ref{SFR_BHacc_31}. 

In Fig. \ref{SFR_BHacc_geo_gas_31} we show the results for three 3:1 mergers
with varying initial gas mass fractions for
a fixed orbit and initial disk geometry (black, green, red lines) and for three 3:1 
mergers with varying initial orbits and orientations for a fixed gas mass fraction (blue lines). 
The initial gas fraction has a large effect on the height of the star
formation and BH accretion peaks, with larger initial gas fractions producing
higher values, as expected. This results also in relatively large differences 
in the final BH masses, with the  $f_{\rm gas}=0.8$ simulations producing
final BH masses that are larger by an order of magnitude compared to the 
$f_{\rm gas}=0.2$ runs. The variation of the orbit and initial geometry
for a fixed gas mass fraction produces much smaller differences. The peaks of the 
star formation rates and BH accretion rates only vary within a factor of two
with changing orbits and initial disk geometries. And although the time of the BH merging varies
somewhat between different geometries the final BH mass is virtually
unchanged. This was already shown to be the case for 1:1 mergers 
by \citet{2005MNRAS.361..776S} and Fig. \ref{SFR_BHacc_geo_gas_31} confirms
the same conclusion for 3:1 mergers. Hence the final BH mass is sensitive to
the initial gas mass fraction of the galaxies, but relatively insensitive to
the initial orbit and orientation of the galaxies. 

Figure \ref{MBH-sigma_31} shows the $M_{\rm BH}-\sigma$ (left panel) 
and $M_{\rm BH}-M_{*}$ (right panel) relations, extracted from 
our total sample of 36 simulated mergers, with the filled triangles and circles depicting
3:1 and 1:1 merger remnants, respectively, together with the observed relations
by \citet{2002ApJ...574..740T} and  \citet{2004ApJ...604L..89H}.
We perform a Levenberg-Marquardt least-squares fit to the $M_{\rm BH}-\sigma$ relation as defined in
Eq. \ref{eq:MBH-sigma}. Table \ref{mbh_sigma_31} lists the best fit values for 
the normalization coefficient $a$ and the slope $b$ together with
their $1 \sigma$ uncertainties derived using the Bootstrap method. In addition we list the
dispersion $\Delta_{\log M_{\rm BH}}$ about each best-fit relation. The
fitting analysis was performed for our total merger sample, for the 3:1 and 1:1
samples separately and for the sample split as a function of the initial gas mass fraction. 
Our best fit normalization coefficient and slopes for the total sample are marginally lower compared to the
observational values of \citet{2002ApJ...574..740T}, but still well within the
error limits. Furthermore the intrinsic $M_{\rm BH}-\sigma$ dispersion induced
primarily by varying the initial gas fraction is in good agreement with the
observed dispersion of $\Delta_{\log M_{\rm BH}}=0.25-0.3$. We also note that our results are
in good agreement with the $z=0$ $M_{\rm BH}-\sigma$ relation derived by 
\citet{2006ApJ...641...90R}. We do not find any statistically significant
difference between the 3:1, 1:1 and the complete merger sample, with all three
samples resulting in virtually the same normalization coefficients and slopes
for the $M_{\rm BH}-\sigma$ relation. The normalization coefficient steadily
increases with increasing initial gas mass fraction. The dispersion within a
sample with constant gas mass fraction is in general low, with the exception
of the sample with the largest gas mass fraction. This can be taken as a further
indication that the dispersion in the $M_{\rm BH}-\sigma$ relation is
primarily driven by changing initial gas mass fractions.
We conclude that the simulated $M_{\rm BH}-\sigma$ relation is overall robust
and in agreement with the observed relation over a wide range of progenitor
masses and initial gas fractions for both equal- and unequal-mass disk mergers.

Similarly, we perform a least-squares fit to the $M_{\rm BH}-M_{*}$ relation as
defined in Eq. \ref{eq:MBH-M*}, with the best fit parameters, parameter errors
and dispersions for our samples listed in Table \ref{mbh_mbul_31}. We here
define $M_*$ as the total mass of all luminous particles within the effective radius
$r_{\rm{e}}$, including disk, bulge and newly formed stellar particles.
This definition assumes that the merger remnant is a pure bulge
system within this radius. We tested this assumption by performing a simple disk-bulge decomposition
following the methodology outlined in \citet{2003ApJ...597...21A}. We
calculated the orbital circularity parameter $\epsilon_{J}=J_{z}/J_{\rm
  circ}(E)$ for all stellar particles within the effective radius, where
$J_{z}$ is the z-component of the specific angular momentum and $J_{\rm
  circ}(E)$ is the specific angular momentum expected for a
corotating circular orbit with binding energy $E$. We found that the fraction of
particles on circular orbits, as defined by $\epsilon_{J}>0.8$ was well below 1\% 
for 1:1 merger remnants and at most 2\% for 3:1 merger remnants.
Thus, for 1:1 and 3:1 mergers the assumption that the remnants are bulge-dominated is valid, whereas for higher
mass ratio mergers this method might induce some additional scatter due to an
underlying surviving disk component in the final merger remnant \citep{2006MNRAS.369..625N}. 
In addition, we found that this definition for $M_*$ correlates reasonably well with total bulge mass as
defined by the stellar bulge particles and finally this definition for $M_*$ is
similar to the one employed by \citet{2008ApJ...676...33D,2007MNRAS.380..877S}
and allows us to directly compare our results to theirs. Thus we adopt the simple definition of the total stellar
mass within the effective radius as a proxy for the total bulge mass for all subsequent measurements of $M_*$. 

The best fit value of the normalization coefficient $(c)$ for our total sample is in excellent
agreement with the observations, but the derived slope $d$ is too steep
compared to the observed sample. The too large value of the slope is a
general problem in all our simulation samples as can be seen in Fig. \ref{MBH-sigma_31}. 
For low-mass stellar systems the simulations typically produce too low BH
masses, whereas for the higher-mass stellar systems the BH mass is slightly
overpredicted, thus producing a tilt in the simulated relation with respect to
the observed relation. The derived dispersion in the simulated relation is
marginally higher than the observed scatter, with the scatter being again
primarily driven by variations in the initial gas mass fraction. Interestingly
the final luminous mass within the effective radius remains virtually constant with changing initial gas
mass fraction. In higher gas mass fraction simulations the lower initial
stellar disk mass is compensated by higher star formation rates resulting in
roughly the same final stellar mass. On the other hand, higher initial gas mass
fractions produce larger final BH masses, thus resulting in vertical scatter
in the $M_{\rm BH}-M_{*}$ relation as can be seen in Fig. \ref{MBH-sigma_31}.   
This also results in a systematic increase in the normalization coefficient
with increasing gas mass fraction, with the slope remaining relatively
constant. A plausible explanation for the discrepancy between the simulated and
observed slopes of the $M_{\rm BH}-M_{*}$ relation lies in the employed
stellar feedback model. A more aggressive stellar feedback including galactic
winds \citep{2003MNRAS.339..289S,2006MNRAS.373.1265O} would primarily affect the low-mass stellar systems, reducing their
final stellar mass, thus lowering the simulated slope and bringing it in
better agreement with the observed slope. Indeed, both the cosmological studies by \citet{2008ApJ...676...33D,2007MNRAS.380..877S}
found a better fit to the expected relations at the low mass end when they
included a model for supernova-driven galactic winds in their simulations.
An alternative explanation for the discrepancy lies in the fact that the progenitors
of lower mass galaxy mergers could have larger initial gas mass fractions
compared to the higher mass merger progenitors. This would also effectively
remove the tilt in the derived correlation bringing it in better agreement with
the observed relation. Given these limitations in the stellar feedback
model employed in our simulations we find that the final fit to the  $M_{\rm
  BH}-M_{*}$ relation is adequate, with a well defined normalization and
a slightly steeper slope compared to the observational relation.

\begin{table}
\caption{Best fit $M_{\rm BH}-M_{*}$ relation for 3:1 and 1:1 mergers}             
\label{mbh_mbul_31}      
\centering          
\begin{tabular}{c c c c c}
\hline\hline       
Sample & N & c & d & $\Delta_{\log M_{\rm BH}}$ \\
\hline                        
Tot sample & 36 & $8.17\pm0.10$ & $1.40\pm0.07$ & 0.44 \\
3:1 sample & 18 & $8.04\pm0.11$ & $1.34\pm0.08$ & 0.47 \\
1:1 sample & 18 & $8.24\pm0.10$ & $1.41\pm0.10$ & 0.38 \\
S1-S2 20\% gas sample & 10 & $7.86\pm0.07$ & $1.34\pm0.05$ & 0.17 \\
S1-S2 40\% gas sample & 10 & $8.28\pm0.08$ & $1.45\pm0.06$ & 0.22 \\
S1-S2 80\% gas sample & 10 & $8.68\pm0.13$ & $1.36\pm0.12$ & 0.29 \\
Observed sample\footnote{From \citet{2004ApJ...604L..89H}.} & 30 & $8.20\pm0.10$ & $1.12\pm0.06$ & $0.30$ \\
\hline                  
\end{tabular}
\end{table}

\subsection{The accretion history of BHs in unequal mergers}

\begin{figure}
\centering 
\includegraphics[width=9cm]{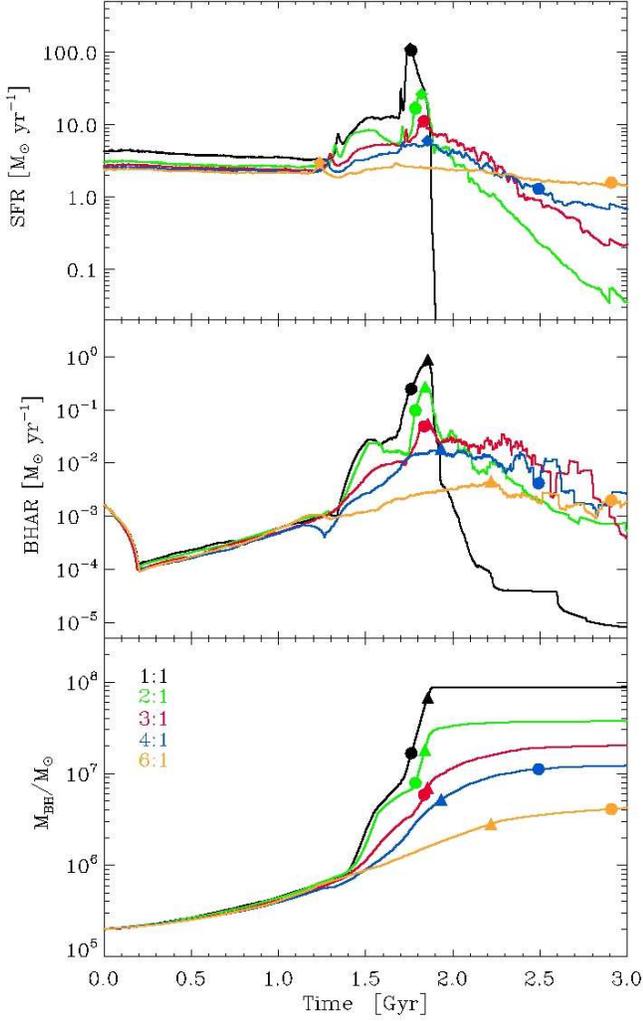}
\caption{The total star formation rate (top), the total black hole accretion rate (middle)
and the evolution of the total black hole mass (bottom) as a function of time for
co-planar prograde (G0) 1:1 (black), 2:1 (green), 3:1 (red), 4:1 (blue) and 6:1
(orange) mergers. The filled circles indicate the time of the BH merger, the
filled diamonds in the top panel and the filled triangles in the bottom two panels
show the location of the maximum star formation and BH accretion peaks, respectively.}
\label{SFR_BHacc_uneq}
\end{figure}

\begin{figure}
\centering 
\includegraphics[width=9cm]{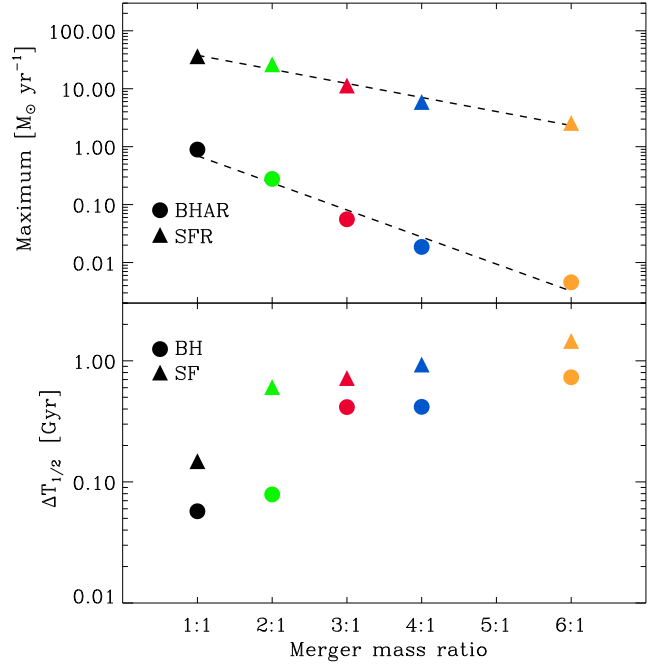}
\caption{The maximum BH accretion and star formation rates and timescales as a
  function of merger mass ratio. The bottom panel shows the half mass growth time of the BHs (circles)
and the stellar mass (triangles) centered on the maximum BH accretion
and star formation rates, respectively. The top panel shows the corresponding
peak BH accretion (circles) and star formation rates (triangles) during the
corresponding half mass growth time  $\Delta{T_{1/2}}$. The dashed line
is the best resulting fit of  Eq. \ref{eq:uneq_fit} to the peak values.}
\label{SFR_BH_acctime_uneq}
\end{figure}

We now proceed to study the BH accretion and star formation histories for
unequal-mass mergers with varying mass ratios. To this end we ran five mergers with
mass ratios of 1:1, 2:1, 3:1, 4:1 and 6:1 on co-planar prograde orbits (G0). This
orbital configuration was chosen in order to maximize the non-axisymmetric torques
during the merger resulting in the most intensive bursts in BH accretion and star  
formation activity \citep{2008MNRAS.tmp...33C}. All the progenitor galaxies had 
initial gas mass fractions of 20\% with the primary progenitor having  $v_{\rm
  vir}=160\ \rm{kms^{-1}}$ at the standard numerical resolution. The
secondary galaxies were scaled down in mass and numerical resolution
accordingly in order to maintain equal mass resolution per particle. The
initial seed BH mass was set to $M_{\rm{BH,init}}=10^{5} M_{\odot}$ for all models.
Again, the initial separation and pericentric distances are set to the mean of
the virial radius and twice the mean of the scale radius respectively as
detailed in \S \ref{merg_pre_mbh_sig}.

In Fig. \ref{SFR_BHacc_uneq} we show the evolution of the resulting star
formation (top), BH accretion (middle) and BH mass (bottom) for the five mergers as a function of
time. Increasing the mass ratio of the merger systematically lowers the
peak star formation rate and increases the duration of star formation activity
after the merger. For the highest mass ratio merger the star formation rate is virtually
constant throughout the simulation with only a mild peak during the first
passage (see also \citealp{2007A&A...468...61D,2007A&A...476.1179B}). 
A similar evolution is seen in the BH accretion rates, with the BH accretion
rates systematically peaking at higher rates with decreasing merger mass ratios. For the higher mass ratio mergers no well defined 
BH accretion peak is discernable, instead the BH accretion rates stay roughly constant
on an elevated level for a prolonged time period compared to the pre-merger BH accretion rates. The
evolution of the BH accretion rates are mirrored in the growth of the BH
masses, with the final BH mass being systematically lower for increasing mass
ratio of the merger. Furthermore, the slope of the $M_{\rm{BH}}$ growth as a
function of time becomes shallower with increasing merger mass ratio. There is
also a systematic delay in the time of the BH merger with increasing mass
ratio, as indicated by the filled circles in Fig. \ref{SFR_BHacc_uneq}. For
the lower mass ratio mergers the peak of the BH activity (the filled triangles
in Fig. \ref{SFR_BHacc_uneq}) typically occurs shortly after the merging of
the BHs, whereas for the higher mass ratio mergers the peak of the BH activity
is not directly related to the merging time of the BHs.  

In Fig. \ref{SFR_BH_acctime_uneq} we study the duration of the BH accretion
and star formation activity as a function of merger mass ratio. We define a
half-mass growth time $\Delta{T_{1/2}}$ during which half of the final BH mass
and half of the total stellar mass is formed respectively. In both cases
$\Delta{T_{1/2}}$ is centered on the peak of the corresponding activity, the
BH accretion on the maximum of the BH accretion rate (the triangles in
Fig. \ref{SFR_BHacc_uneq}) and the star formation rate on the peak of the SFR
marked with diamonds in Fig. \ref{SFR_BHacc_uneq}. The resulting half-mass
timescales are shown in the bottom panel of Fig. \ref{SFR_BH_acctime_uneq}.  
The $\Delta{T_{1/2}}_{\rm BH}$ is strongly correlated with the mass
ratio of the merger. For 1:1 and 2:1 mergers the growth of the $M_{\rm{BH}}$
is very concentrated in time, with half of the final BH mass growth occurring in less
than 100 Myr. For the 3:1 and 4:1 mergers $\Delta{T_{1/2}}_{\rm BH}\sim 0.5 \ \rm
Gyr$, with the 6:1 merger resulting in $\Delta{T_{1/2}}_{\rm BH}\sim 1 \ \rm
Gyr$. In the
lower mass ratio mergers a significant fraction of the gas is funnelled to the
center during the merger, enabling the BHs to experience rapid Eddington limited growth
(Eq. \ref{Eddington}). For larger mass ratio mergers the tidally driven non-axisymmetric torques
are weaker and a lower fraction of the gas reaches the center resulting in
lower BH accretion rates and longer BH half-mass growth timescales. 

The corresponding stellar half-mass timescales $\Delta{T_{1/2}}_{\rm star}$ (triangles in
the bottom panel Fig. \ref{SFR_BH_acctime_uneq}) also show a clear correlation with the mass ratio of the
merger. In the 1:1 merger half of the final stellar mass is formed in a short
major burst lasting about $\Delta{T_{1/2}}_{\rm star}\sim 0.15 \ \rm Gyr$. This
value is comparable to the star formation timescale derived 
by \citet{2008MNRAS.tmp...33C}, who calculated a full width at half maximum of
$\rm{FWHM}\sim 0.1 \ \rm{Gyr}$ for the star formation peak of a typical 1:1 merger. 
For higher mass-ratio mergers the resulting star formation
timescales are $\Delta{T_{1/2}}_{\rm star}\sim 0.7-1.0 \ \rm{Gyr}$, with the highest
mass-ratio merger having the longest timescale of $\Delta{T_{1/2}}_{\rm
  star}\sim 1.5 \ \rm{Gyr}$. In these cases it is not straightforward to
relate our $\Delta{T_{1/2}}_{\rm star}$ to the burst timescales defined by
\citet{2008MNRAS.tmp...33C}, as the higher mass ratio mergers have less well
defined star formation peaks with the star formation distributed more evenly 
over time. However, the general trend that increasing the merger mass ratio
results in increasingly longer star formation timescales agrees well with the 
results of \citet{2008MNRAS.tmp...33C}. 

Finally we plot in the top panel of Fig. \ref{SFR_BH_acctime_uneq} the corresponding
maximum black hole accretion and star formation rates during their respective
half mass growth timescales. By defining the variable $q$ as the mass ratio
between the primary and secondary component we can fit the logarithms of the
peak BH accretion and star formation rates with the following linear relation

\begin{equation}
\log \rm{Maximum} \ [M_{\odot} \rm{yr^{-1}}]=a_{0}+a_{1} \times q,
\label{eq:uneq_fit}
\end{equation}
where $a_{0}$ and $a_{1}$ are the inferred normalization and slope respectively. 
Both the maximum BH accretion rates and star formation rates are well fitted
by Eq. \ref{eq:uneq_fit} (dotted lines in Fig. \ref{SFR_BH_acctime_uneq}) 
resulting in the following best fitting parameters respectively: 
$(a_{0,BHAR}=0.31, a_{1,BHAR}=-0.47)$ 
$(a_{0,SFR}=1.82, a_{1,SFR}=-0.24)$. The ratio of the peak BH accretion rate
to the peak star formation rate is thus of the order of $\dot{M}_{\rm
  BH,peak}/\dot{M}_{\rm SF,peak} \sim 10^{-2}$. On the other hand, the ratio of
the mean BH accretion rate to the mean star formation averaged over the whole
simulation is closer to $\dot{M}_{\rm BH,mean}/\dot{M}_{\rm SF,mean} \sim
10^{-3}$, with this ratio varying systematically between  
$\dot{M}_{\rm BH,mean}/\dot{M}_{\rm SF,mean}=2\cdot10^{-3}$ (1:1 merger) and 
$\dot{M}_{\rm BH,mean}/\dot{M}_{\rm SF,mean}=0.2\cdot10^{-3}$ (6:1
merger). Interestingly, recent observations of $z\sim2$ galaxies by \citet{2007ApJ...670..156D,2007ApJ...670..173D}
also find indications of an universal ratio of $\dot{M}_{\rm BH}/\dot{M}_{\rm SF} \sim
10^{-3}$ and this is also expected from the observed $M_{{\rm BH}}-M_{*}$
relation. Naturally the normalizations and to some extent the slopes of the derived BH and SFR activity relations are
dependent on the initial gas mass fraction and orbital geometry of the
system. However, the derived ratios between the BH accretion and SF rates
should be valid and we find in our simulations that for any given orbital configuration and initial
gas mass fraction the ratio of the peak BH accretion and SF rates is $\sim
10^{-2}$, whereas the corresponding mean value is of the order of $\sim 10^{-3}$.

\subsection{Mixed E-Sp mergers with BHs}

\begin{table*}
\caption{Mixed E-Sp mergers}             
\centering  
\scriptsize{
\label{Mix_sims}      
\begin{tabular}{c c c c c c c c c c c}
\hline\hline       
                      
Models & Progenitor 1 &  $v_{\rm vir,prog2}$\footnote{All virial velocities in $\rm{kms^{-1}}$} & 
Mass ratio & Orbit & $f_{\rm gas,init}$ 
& $N_{\rm{disk,tot}}$  & $N_{\rm{bulge,tot}}$  &
  $N_{\rm{DM,tot}}$ & $N_{\rm{sim}}$ \\
\hline                    

S1-E-Sp  & S1-1:1 & 50, 80, 160, 320, 500 & 2:1 & G0 & 0.2 & 60000 & 30000 & 90000 & 
  5 \\  
S2-E-Sp & S2-3:1 & 35, 55, 111, 222, 347  & 4:1 & G0 & 0.2 &  33334 & 16666  & 50000 &
  5 \\  
G1-E-Sp  & G1-1:1 & 160 & 2:1 & G7, G10, G13 & 0.2 & 60000 & 30000 & 90000 & 
  3 \\  
G2-E-Sp & G2-3:1 & 111 & 4:1 & G4, G10, G14 & 0.2 &  33334 & 16666  & 50000 &
  3 \\  

\hline                  
\end{tabular}
}
\end{table*}

\begin{figure}
\centering 
\includegraphics[width=9cm]{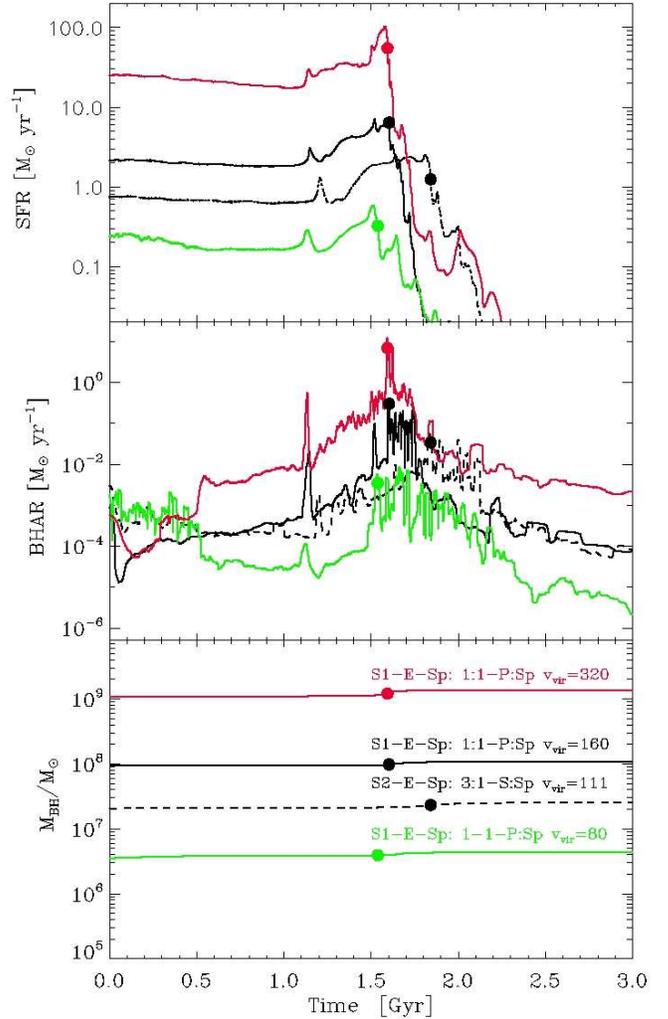}
\caption{The total star formation rate (top), the total black hole accretion rate (middle)
and the evolution of the total black hole mass (bottom) as a function of time for
four E-Sp mixed mergers. The solid lines show mixed E-Sp mergers of 1:1 merger remnants
  with primary spiral disks and the dashed line shows
  a mixed E-Sp merger between a 3:1 merger remnant and a secondary spiral
  disk. The gas disks have initially $f_{\rm gas}=20\%$ and the virial
  velocities as indicated in the Figure.  The filled circles
indicate the time of merging of the BHs.}  
\label{SFR_BHacc_mix}
\end{figure}

\begin{figure*}
\centering 
\includegraphics[width=18cm]{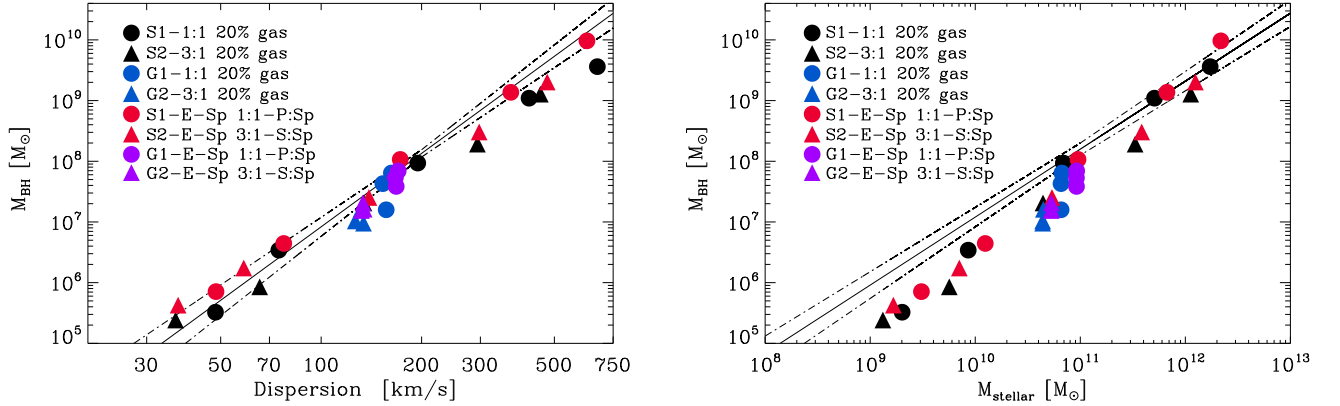}
\caption{The $M_{\rm BH}-\sigma$ (left) and $M_{{\rm BH}}-M_{*}$ (right) relations for our sample of mixed E-Sp mergers. The
  merger remnants of the original 1:1 and 3:1 mergers are shown by black
  and blue filled circles and triangles for the S and G merger series, 
  respectively (Table \ref{sims}). The mixed merging of 1:1 merger remnants
  with primary spiral disks with  $f_{\rm gas}=20\%$ (1:1-P:Sp) are shown with filled red
  (S1-series) and purple (G1-series) circles. The mixed merging of 3:1 merger
  remnants with secondary spiral disks with $f_{\rm gas}=20\%$ (3:1-S:Sp) are shown with filled red
  (S2-series) and purple (G2-series) triangles. The lines
  show the observed relations with errors by \citet{2002ApJ...574..740T} (left
  panel) and \citet{2004ApJ...604L..89H} (right panel), respectively.}
\label{MBH-sigma_mix}
\end{figure*}

We now turn our attention to mixed E-Sp mergers, in which a gas-poor
early-type galaxy merges with a gas-rich spiral disk galaxy. To this end we 
simulated to our knowledge for the first time a large sample of mixed
E-Sp mergers. For early-type progenitors we use the 
merger remnants of 1:1 and 3:1 mergers discussed in \S \ref{uneq_eq}, whereas
the progenitor disk galaxies are setup using the disk models discussed in \S \ref{galaxy_models}.
Our mixed merger sample consisted of five 1:1 merger remnants that are merged with the
primary disks of the original simulations (S1-E-Sp) and another five runs in which 3:1
merger remnants are merged with the corresponding secondary disks of the
original simulations (S2-E-Sp). This sample of ten mixed mergers were run on the same G0
orbit as the original mergers. In addition, we simulated six E-Sp mergers of
1:1 and 3:1 merger remnants with primary and secondary disks respectively
(G1-E-Sp and G2-E-Sp) with varying initial orientations
and orbital geometries as indicated in Table \ref{Mix_sims}. This results in a 
total sample of 16 mixed E-Sp mergers. The initial gas
fraction was set to 20\% in both the original simulations that produced the
E-progenitors and in the disk progenitors. The black hole mass of the early-type
progenitors is set by the result in the original simulation and is typically  $M_{\rm BH}\sim
10^{6}-10^{9} M_{\odot}$, whereas the initial seed BH mass of the disk galaxies is
always fixed at $M_{\rm BH}= 10^{5} M_{\odot}$.
 
We plot the corresponding total star formation rates (top), the total BH accretion rates
(middle) and the evolution of the total BH mass (bottom) in Fig. \ref{SFR_BHacc_mix}
for a sample of four mixed E-Sp mergers. The three solid lines show mergers between
1:1 merger remnants and primary disks and the dashed line a merger between a
3:1 merger remnant and a secondary disk, with the color scaling indicating
the virial velocity of the disk progenitors. The combined star formation rate of the mixed E-Sp mergers
is lower than in Sp-Sp (Fig. \ref{SFR_BHacc_31}) mergers, due to the lower
amount of cold gas available for star formation. The disk progenitor contains 
$f_{\rm gas}=20\%$ of gas initially, whereas the early-type progenitors 
typically have an initial gas mass fraction of $f_{\rm gas}\sim5\%$, with 
typically only $\sim1\%$ of this gas being cold and dense.
After the merger of the E-Sp galaxies the
star formation rate declines rapidly in all the merger remnants.
The black hole accretion rates are somewhat lower than in Sp-Sp mergers, but
still relatively high due to the large amount of available gas in the central
regions of the merger remnants originating from the disk progenitors. The BH accretion rates show a clear peak,
both during the time of the first passage and at the time of final
coalescence. We define a BH mass growth factor as $f_{\rm
  BH,insitu}=\Delta M_{\rm BH,insitu}/M_{\rm BH,final}$, which gives the ratio of
BH mass growth due to gas accretion during the simulation with respect to the final BH mass.
Quantitatively, the fraction of the BH mass that accumulates by gas accretion during the mixed
mergers is in the range of $f_{\rm BH,insitu}\sim20-50\%$, with typical mean values 
of $f_{\rm BH,insitu}\sim30\%$. Thus in most cases the majority of the final
BH mass is still contributed by the massive seed BHs residing in the early-type progenitors. 

\begin{table}
\caption{Best fit $M_{\rm BH}-\sigma$ relation for E-E and E-Sp mergers}             
\label{mbh_sigma_rem}      
\centering          
\begin{tabular}{c c c c c}
\hline\hline       
                      
Sample & N & $a$ & $b$ & $\Delta_{\log M_{\rm BH}}$ \\
\hline                    
Progenitor sample & 16 &  $7.83\pm0.04$ & $3.53\pm0.11$ & 0.16 \\
E-Sp Mixed sample & 16 & $8.03\pm0.04$ & $3.55\pm0.12$ & 0.13 \\
E-E Remerger sample & 16 & $8.13\pm0.05$ & $3.41\pm0.10$ & 0.18 \\

\hline                  
\end{tabular}
\end{table}

The $M_{\rm BH}-\sigma$ (left panel) and $M_{{\rm BH}}-M_{*}$ (right panel) relations for our mixed
E-Sp merger (red and purple symbols) sample together with the E-type progenitor
population (black and blue symbols) is plotted in
Fig. \ref{MBH-sigma_mix}.  All the progenitor disk galaxies have initial seed
BH masses of $M_{\rm BH}= 10^{5} M_{\odot}$ with the velocity dispersion $\sigma$
depending on the bulge mass of the progenitor galaxy. Thus the BHs of the disk
galaxies are not initially on the $M_{\rm BH}-\sigma$ relation and they are
also not plotted in Fig. \ref{MBH-sigma_mix}. Using a least-squares fit and
Bootstrap technique we
derive the best fitting parameters for the $M_{\rm BH}-\sigma$ relation
of our mixed merger and progenitor samples (see Table \ref{mbh_sigma_rem}). For the mixed
mergers the normalization parameter $a$
is 0.2 dex higher than for the Sp-Sp progenitor sample, but 0.1 dex lower than for
the E-E mergers. The combined in-situ growth due to gas accretion and merging of the BHs
thus result in final BH masses that are very similar to the ones derived in
both the Sp-Sp (Fig. \ref{MBH-sigma_31}) and E-E mergers (Fig. \ref{MBH-sigma_rem}). The mixed merging of E-Sp
galaxies seems to maintain the slope $b$ of the $M_{\rm BH}-\sigma$ relation
compared to the progenitor sample and even slightly reduces the dispersion.
The mixed mergers primarily increase the final BH mass, with smaller changes in
the final velocity dispersion compared to the initial dispersion of the early-type
progenitors. Overall the slope and dispersion are lower than the observed
\citet{2002ApJ...574..740T} values of $b=4.02\pm0.32$, $\Delta_{\log M_{\rm
    BH}}=0.25-0.3$. This is primarily due
to the fact that all of the original mergers and disk progenitors had initial gas mass
fractions of 20\%. Including mixed mergers of merger remnants and disks with higher initial
gas mass fractions would increase the normalization, slope and scatter in the  
derived $M_{\rm BH}-\sigma$ relation (see Table \ref{mbh_sigma_31}).
The mixed mergers undergo a similar phase of strong self-regulated BH growth as
the Sp-Sp mergers, allowing the BH feedback to regulate the final BH mass
in agreement of the $M_{\rm BH}-\sigma$ relation. Thus we conclude that the $M_{\rm BH}-\sigma$ relation should remain
relatively stable even after several generations of mixed E-Sp mergers.

Finally, we also perform a least-squares fit to the $M_{\rm BH}-M_{*}$
(Eq. \ref{eq:MBH-M*}) relation for our mixed E-Sp sample with the best fit
parameters listed in Table \ref{mbh_m*_rem}. The mixed merging of an E-Sp
galaxy population effectively maintains the original $M_{\rm BH}-M_{*}$
relation of the progenitor sample, even slightly reducing the intrinsic dispersion in the relation.
The self-regulated stellar and BH feedback models regulate the star formation
process and BH mass growth in mixed mergers similarly to Sp-Sp mergers
thus ensuring that the derived original $M_{\rm BH}-M_{*}$ relation is
maintained. We thus conclude that even repeated generations of mixed E-Sp
mixed mergers should not greatly affect the observed $M_{\rm BH}-M_{*}$ relation.

\begin{table}
\caption{Best fit $M_{\rm BH}-M_{*}$ relation for E-Sp and E-E mergers}             
\label{mbh_m*_rem}      
\centering          
\begin{tabular}{c c c c c}
\hline\hline       
                      
Sample & N & c & d & $\Delta_{\log M_{\rm BH}}$ \\
\hline                    
Progenitor sample & 16 &  $7.78\pm0.07$ & $1.35\pm0.05$ & 0.21 \\
E-Sp Mixed sample & 16 & $7.83\pm0.05$ & $1.39\pm0.07$ & 0.16 \\
E-E Remerger sample & 16 & $7.86\pm0.05$ & $1.38\pm0.04$ & 0.17 \\

\hline                  
\end{tabular}
\end{table}

\subsection{E-E remergers with BHs}
\label{sec:E-E}

\begin{table*}
\caption{E-E remergers}             
\centering 
\scriptsize{
\label{EE_sims}      
\begin{tabular}{c c c c c c c c c c c}
\hline\hline       
                      
Models & Progenitor 1 &  Progenitor 2 & 
Mass ratio & Orbit & $f_{\rm gas,init}$ 
& $N_{\rm{disk,tot}}$  & $N_{\rm{bulge,tot}}$  &
  $N_{\rm{DM,tot}}$ & $N_{\rm{sim}}$ \\
\hline                    

S1-E-E  & S1-1:1 & S1-1:1 & 1:1 & G0 & 0.2 & 80000 & 40000 & 120000 & 
  5 \\  
S2-E-E & S1-1:1 & S2-3:1 & 3:2 & G0 & 0.2 &  66667 & 33333  & 100000 &
  5 \\  
G1-E-E  & G1-1:1 & G1-1:1 & 1:1 & G7, G10, G13 & 0.2 & 80000 & 40000 & 120000 & 
  3 \\  
G2-E-E & G2-3:1 & G2-3:1 & 1:1 & G4, G10, G14 & 0.2 &  26666 & 53334  & 80000 &
  3 \\  

\hline                  
\end{tabular}
}
\end{table*}

\begin{figure}
\centering 
\includegraphics[width=9cm]{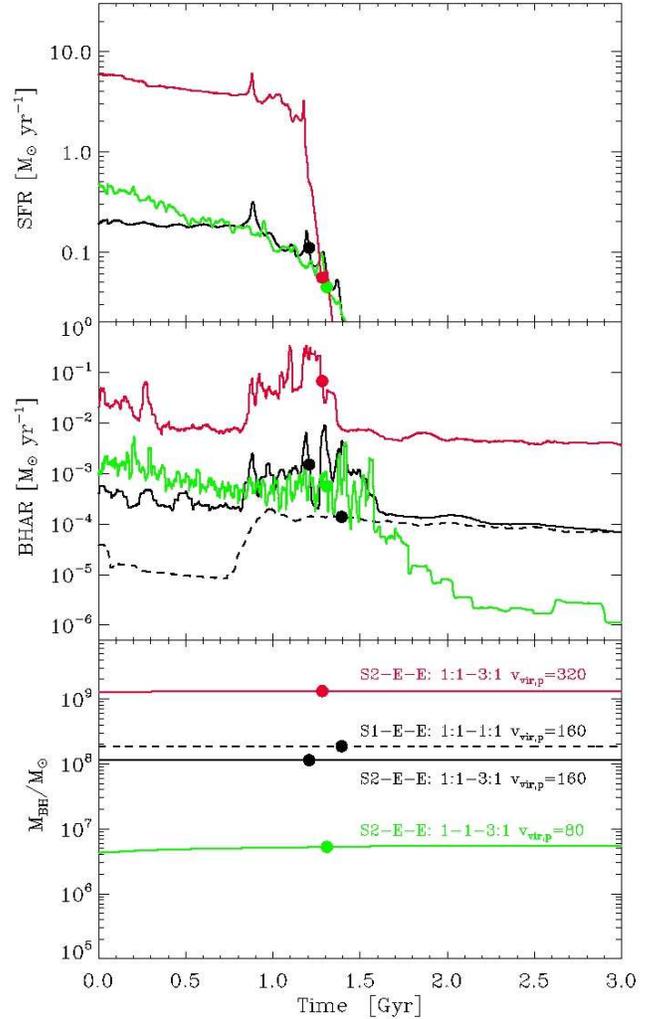}
\caption{The total star formation rate (top), the black hole accretion rate (middle)
and the evolution of the total black hole mass (bottom) as a function of time for
four E-E remergers. The solid lines show remergers between 1:1 and
3:1 merger remnants and the dashed line shows a E-E remerger between two 1:1
merger remnants. The virial velocities of the primary galaxies in the original
simulations are as indicated in the Figure. The filled circles
indicate the time of merging of the BHs.} 
\label{SFR_BHacc_rem}
\end{figure}

\begin{figure*}
\centering 
\includegraphics[width=18cm]{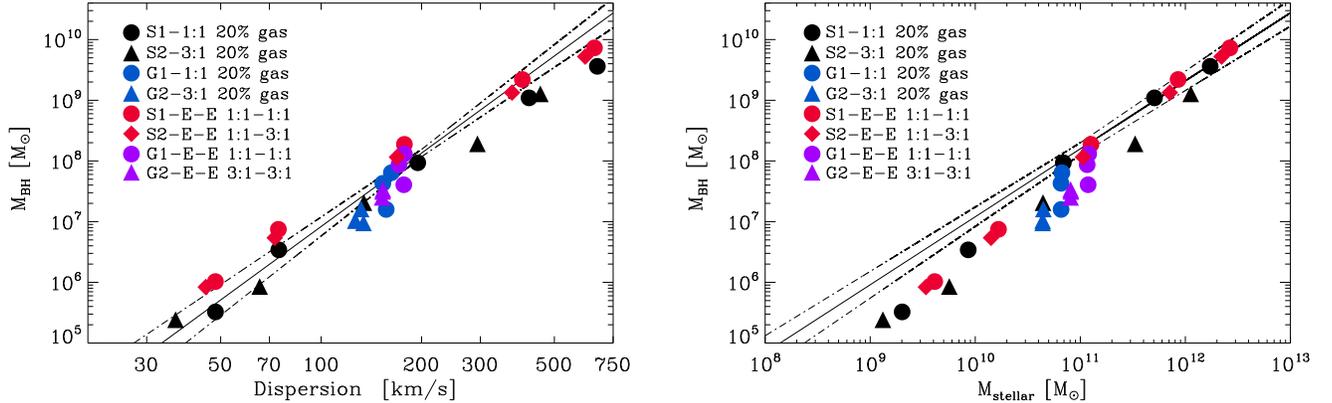}
\caption{The $M_{\rm BH}-\sigma$ (left) and $M_{{\rm BH}}-M_{*}$ (right) relations for our sample of E-E remergers. The
  merger remnants of the original 1:1 and 3:1 mergers are shown by black
  and blue filled circles and triangles for the S and G merger series,
  respectively (Table \ref{sims}). The remerging of 1:1-1:1 merger remnants are
  shown with filled red (S1-series) and purple (G1-series) circles, the remerging of 1:1-3:1 merger
  remnants (S2-series) are shown with red diamond symbols and finally the remerging of
  3:1-3:1 remnants (G2-series) by purple triangles. The lines
  show the observed relations with errors by \citet{2002ApJ...574..740T} (left
  panel) and \citet{2004ApJ...604L..89H} (right panel), respectively.} 
\label{MBH-sigma_rem}
\end{figure*}

Finally, we studied E-E
mergers including BH feedback by remerging a sample of our merger
remnants produced by the 3:1 and 1:1 mergers discussed in \S \ref{uneq_eq}.  
We set up and ran five equal-mass remergers of our S1-1:1 (Table \ref{sims}) merger remnants (S1-E-E), with the
initial progenitor disks of the original simulations having virial velocities
in the range $v_{\rm vir}=50-500 \ \rm{km s^{-1}}$. In addition, we remerged five of our 1:1 merger remnants with
the corresponding set of 3:1 merger remnants (S2-E-E), resulting in a mass ratio of 3:2
(see Table \ref{EE_sims}). All of the mergers were run on identical orbits to
the original mergers and the initial gas fraction of the progenitors for the
original simulations was fixed at 20\%. In addition to the ten S-series remergers that were run on
the G0 orbit, we simulated three E-E mergers of 1:1 merger remnants (G1-E-E) and three
E-E mergers of 3:1 merger remnants (G2-E-E) with varying initial orientations
and orbital geometries as indicated in Table \ref{EE_sims}, thus resulting in a total sample
of 16 E-E remergers.

In Fig. \ref{SFR_BHacc_rem} we plot the star formation rate (top), the BH
accretion rate (middle) and the evolution of the BH mass (bottom) for three
1:1-3:1 E-E remergers (solid lines) and one 1:1-1:1 E-E remerger (dashed line).
Here the coloring indicates the virial velocities of the primary galaxies in
the original simulations, $v_{\rm vir,p}=80 \ \rm{kms^{-1}}$ (green),  
$v_{\rm vir,p}=160 \ \rm{kms^{-1}}$ (black) and $v_{\rm vir,p}=320
\ \rm{kms^{-1}}$ (red). 
The initial star formation rates are generally very low due to the
low gas fractions of $f_{\rm gas}\sim 1-5 \%$ and $f_{\rm gas}\sim 5-10 \%$
for the 1:1 and 3:1 merger remnant E-progenitors, respectively. The initial
gas fraction directly depends on the strength of the initial interaction that gave rise to
the merger remnants used as progenitors for the E-E remergers. Increasing the
masses of the progenitors and using more direct planar orbits (G0) produces
more violent initial encounters, thus decreasing $f_{\rm gas}$.
In addition, most of the gas is hot with low densities and the fraction of cold star-forming
gas is typically below $\lesssim 1 \%$ of the total gas mass.
As can be seen in Fig. \ref{SFR_BHacc_rem} the star formation is very effectively terminated 
shortly after the merging of the E-E progenitors on comparable timescales to
1:1 Sp-Sp mergers and thus more efficiently than in 3:1 Sp-Sp and mixed E-Sp mergers.
For the 1:1-1:1 E-E remerger
the star formation rate is zero throughout the simulation as the star
formation was already completely terminated in the original 1:1 merger that
served as a progenitor galaxy for the E-E merger (see
Figs. \ref{SFR_BHacc_31} and \ref{SFR_BHacc_uneq}). 

The black hole accretion rates strongly
correlate with the available amount of gas in the central regions and are thus
typically one to two orders of magnitude lower for dry E-E mergers compared to the
mixed E-Sp and Sp-Sp mergers. This is also reflected in the total BH mass
growth plots (Fig. \ref{SFR_BHacc_rem}) that remain virtually flat throughout the simulations. 
The growth of the final BH mass is primarily due to the merging of the BHs with the in-situ growth of
the BH mass through gas accretion being relatively low. 
The exact fraction of BH mass grown in-situ ($f_{\rm BH,insitu}$) due to gas accretion during the
simulation with respect to the final BH mass, depends on the amount of
available gas initially. 
For remerging of 1:1 merger remnants
this fraction is typically very low at $f_{\rm BH,insitu}<1\%$, whereas for the 
remerging of  1:1-3:1 merger remnants the fraction is higher at $f_{\rm
  BH,insitu}\sim10\%$, with the highest fraction of $f_{\rm
  BH,insitu}\sim20\%$ produced for remerging of 3:1-3:1 merger remnants, as
expected. There is relatively large scatter in the $f_{\rm BH,insitu}$
fraction as a function of the orbital configuration and the initial masses of
the progenitor galaxies primarily set by the corresponding scatter in the
initial gas mass fractions of the progenitor galaxies. 
All the fractions above are estimates for only first generation remergers, with the
$f_{\rm BH,insitu}$ ratio rapidly decreasing with increasing generations of remergers.
The same is true for the star formation, E-E remerging effectively quenches
the star formation (Fig. \ref{SFR_BHacc_rem}) and already a second generation of dry merging is
expected to show very weak, if any, signatures of residual star formation.  

In Fig. \ref{MBH-sigma_rem} we plot the $M_{\rm BH}-\sigma$ (left panel) and 
the $M_{\rm BH}-M_{*}$ (right panel) relations for our
E-E remerger (red and purple symbols) sample together with the progenitor
population (black and blue symbols). The best fitted $M_{\rm BH}-\sigma$
parameters together with the $1\sigma$ uncertainties for our E-E remerger
sample are listed in Table \ref{mbh_sigma_rem}.
The normalization parameter $a$
is 0.3 dex higher for the E-E remerger sample compared to the progenitor
sample and 0.1 dex higher compared to the mixed E-Sp sample. However  
the slope $b$ is marginally shallower compared to both the progenitor and
mixed merger samples. The E-E remerging of
the merger remnants increases the final BH mass, but leaves the velocity
dispersion more or less unchanged, resulting in an increase of $a$. 
The relative increase in $M_{\rm BH}$ is slightly larger for low-mass systems
compared to higher-mass systems, resulting in a slightly lower slope $b$. 
This is especially true for the remergers on the G0 orbit,
whereas remergers on other orbits (the purple G-series in Fig. \ref{MBH-sigma_rem}) 
also increase slightly the velocity dispersion of the resulting merger remnants (see
also \citealp{2006ApJ...641...21R}). The overall lower value of the slope $b$ is again due to the
fact that only E-E mergers of progenitors with initial gas fractions of
$f_{\rm gas}=20\%$ in the original simulations were included in constructing
the E-E sample. We thus conclude that dry E-E remerging typically increases the normalization
parameter $a$ and lowers the slope $b$ slightly, but a single generation of dry
remerging is not able to destroy the $M_{\rm BH}-\sigma$ relation initially generated by
gas-rich mergers (see also \citealp{2006ApJ...641...21R}). E-E remerging also
increases slightly the dispersion in the fitted relation, but the variation in
the initial gas fraction still remains the dominant source of
scatter. Unlike Sp-Sp and mixed E-Sp mergers the gas-poor E-E mergers only
experience a weak self-regulatory BH growth phase, with this phase missing
altogether for the gas-poorest E-E mergers. This lack of the self-regulating feedback phase can
thus explain some of the larger scatter in the final E-E remerger $M_{\rm BH}-\sigma$ relation. 

\begin{figure*}
\centering 
\includegraphics[width=18cm]{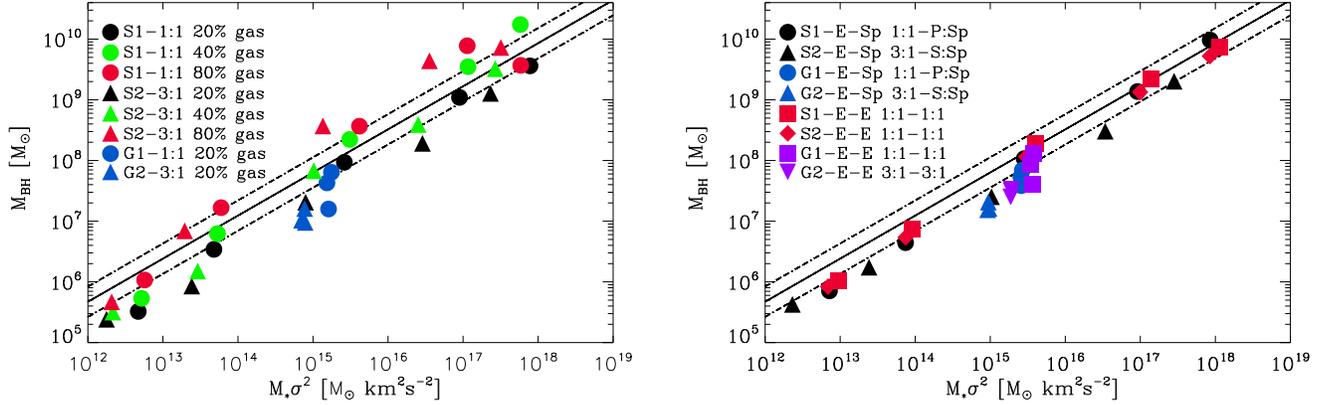}
\caption{The binding energy correlation, which is statistically equivalent to
  the BHFP. In the left panel we plot the correlation for our complete sample
  of 3:1 (triangles) and 1:1 (circles) mergers, with the colors indicating
  variations in the initial gas mass fractions (black, green, red) and in the
  orbital geometries (blue). In the right panel we plot the
  corresponding relation for our mixed E-Sp (circles and triangles) and the
  E-E remerger sample (squares, diamonds and downward pointing triangles). 
  The lines show the observed relation with scatter from the compilation of 
  \citet{2007ApJ...669...45H}}.
 \label{MBH-Ebind}
\end{figure*}

We also perform a least-squares fit to the $M_{\rm BH}-M_{*}$
(Eq. \ref{eq:MBH-M*}) relation for both the E-E remerger sample and the
progenitor sample with  
the best fit parameters, parameter errors and dispersions listed in Table \ref{mbh_m*_rem}.
Dry E-E remerging effectively maintains the normalization coefficient $(c)$ and slope
$(d)$ of the progenitor sample, with a very minor increase of both parameters that lie
within the $1\sigma$ uncertainties of the corresponding progenitor
parameters. Furthermore dry remerging roughly maintains the dispersion derived
for the progenitor population. This is expected since the in-situ growth from gas of both the BH
and stellar mass are very low during the simulation. For the limiting case of no available gas
and no in-situ growth of the BH and stellar mass, the original 
$M_{\rm BH}-M_{*}$ relation would be perfectly conserved. Thus we expect that
increasing generations of E-E mergers would not substantially increase the
scatter in the observed $M_{\rm BH}-M_{*}$ relation. The discrepancy between
the derived $M_{\rm BH}-M_{*}$ parameters for the E-E remergers and the observed parameters
would be remedied by remerging merger remnants with higher initial gas mass
fractions (increases $c$) and by employing more effective stellar feedback
models (lowers $d$) as was already discussed in \S \ref{uneq_eq}.

\subsection{The Black Hole fundamental plane}

\begin{table}
\caption{Best fit binding energy relation for 3:1, 1:1,  E-Sp and E-E mergers}             
\label{Ebind_tab}      
\centering          
\begin{tabular}{c c c c c}
\hline\hline       
                      
Sample & N & e & f & $\Delta_{\log M_{\rm BH}}$ \\
\hline                    
3:1 sample & 18 & $8.05\pm0.10$ & $0.79\pm0.04$ & 0.41 \\
1:1 sample & 18 & $8.15\pm0.08$ & $0.81\pm0.05$ & 0.32 \\
S1-S2 20\% gas sample & 10 & $7.85\pm0.05$ & $0.76\pm0.03$ & 0.14 \\
S1-S2 40\% gas sample & 10 & $8.22\pm0.06$ & $0.84\pm0.03$ & 0.17 \\
S1-S2 80\% gas sample & 10 & $8.54\pm0.12$ & $0.79\pm0.06$ & 0.28 \\
E-Sp Mixed sample & 16 & $7.91\pm0.06$ & $0.77\pm0.03$ & 0.15 \\
E-E Remerger sample & 16 & $7.98\pm0.05$ & $0.76\pm0.02$ & 0.18 \\
Total sample & 68 & $8.04\pm0.05$ & $0.79\pm0.02$ & 0.31 \\
Observed sample\footnote{From \citet{2007ApJ...669...45H}} & 38 & $8.23\pm0.06$ & $0.71\pm0.06$ & 0.25 \\
\hline                 
\end{tabular}
\end{table}

Recent studies by \citet{2007ApJ...669...45H,2007ApJ...669...67H}
have shown that the well established $M_{\rm BH}-\sigma$ and $M_{\rm
  BH}-M_{*}$ relations are only two projections of a more fundamental
  underlying relation. This so called Black Hole Fundamental plane (BHFP)
is defined as $M_{\rm BH}\sim \sigma^{3.0\pm0.3} R^{0.43\pm0.19}$ or 
$M_{\rm BH}\sim M_{*}^{0.54\pm0.17} \sigma^{2.2\pm0.5}$ and is analogous to
  the fundamental plane derived for elliptical galaxies \citep{2007ApJ...669...45H,2007ApJ...669...67H}.
A statistically equivalent formulation of the BHFP is given by the binding
  energy correlation, which relates the total binding energy of the bulge
  $(E\sim M_{*}\sigma^{2})$ to the mass of the the BH. Recent observations by 
\citet{2007ApJ...665..120A,2007ApJ...662L..67B} have shown that the bulge
  gravitational binding energy as traced by the stellar light profile
  correlates strongly with the BH mass. 

Figure \ref{MBH-Ebind} shows the binding energy correlation for our 3:1 and
1:1 disk merger sample (left panel) and for the mixed E-Sp and dry E-E mergers
(right panel), respectively, together with the observed relation from the
compliation of \citet{2007ApJ...669...45H}. We perform a least-squares fit to
the data using the equation below

\begin{equation}
\log (M_{\rm BH}/M_{\odot})=e+f \log (M_*\sigma^2/(M_{0}\sigma_{0}^{2})),
\label{eq:Ebind}
\end{equation}
where $e$ and $f$ are the normalization coefficient and slope, respectively, and
$M_{0}=10^{11} M_{\odot}$ and $\sigma_{0}=200 \ \rm{km/s}$ are the normalization
values. The best values of $e$ and $f$, together with the dispersion for the various
subsamples are given in Tab. \ref{Ebind_tab}. We do not find any statistically
significant difference between the 3:1 and 1:1 merger samples. For both samples the
derived normalizations and slopes are very similar and in good agreement with
the observed relation. Varying the initial gas mass fraction produces more
pronounced differences, with a systematically higher normalization coefficient
for increasing initial gas mass fractions. However, the slope remains constant
within the error bars with changing initial gas mass fractions. The mixed E-Sp
and dry E-E merger samples generally maintain the slope of the progenitor sample, with a
slight increase in the normalization. Finally, combining all the merger samples
together produces a final relation in reasonably good agreement with the 
observed relation and with a similar dispersion to the observed sample. The
tightness of the derived correlation indicates that the final BH mass in all the
merger samples is primarily set by the depth of the gravitational potential well.

We conclude that the bulge binding energy relation provides a robust predictor
for the BH mass over a wide range of progenitor masses and initial gas
fractions for both unequal- and equal-mass disk mergers, mixed E-Sp mergers and E-E mergers.
The dispersion in the bulge binding energy correlation is similar in size to
the dispersion in the $M_{\rm BH}-\sigma$ relation and both these methods
should be preferred over the $M_{\rm BH}-M_{*}$ relation that produces
significantly larger scatter (see \citealp{2007ApJ...665..120A} for a discussion).

\section{Summary and Discussion}
\label{conc}

In this paper we have studied galaxy mergers with BH feedback starting with equal- and unequal-mass disk progenitors, mixed
disk-elliptical progenitors and elliptical-elliptical progenitors. Extending
on detailed previous studies of equal-mass disk mergers
(e.g. \citealp{2006ApJ...645..986R}) our merger sample includes 18 unequal-mass disk mergers, 16 mixed
E-Sp mergers and 16 dry E-E mergers in addition to 18 equal-mass mergers
resulting in a total sample of 68 mergers. The simulations
include radiative cooling, star formation and black holes with their
associated feedback processes as described in \citet{2005MNRAS.361..776S}.

We study the termination of star formation in merger
simulations including BH feedback. We find that the termination of star
formation by BH feedback is significantly less important for unequal-mass
disk mergers compared to equal-mass disk mergers \citep{2005ApJ...620L..79S}.
The timescale for star formation termination systematically increases with
increasing progenitor mass ratios. Similarly, a systematic increase is seen in
the half-mass growth timescales of the BHs, with this timescale varying from
$\sim0.1$ Gyr for equal-mass mergers to $\sim 1$ Gyr for 6:1 mergers.
This systematic trend can be used 
as input in modeling BH accretion more realistically in semi-analytic galaxy
formation models (e.g. \citealp{2006MNRAS.365...11C}).
For mass-ratios of 3:1 and higher mergers with BH feedback are unable to
completely quench the star formation, with the merger remnants showing star
formation rates roughly on the pre-merger level even 1 Gyr after completion of the merger. 

In mixed E-Sp mergers and dry E-E mergers the star formation and BH accretion
rates are in general low due to the low cold gas mass fraction.
The star formation is efficiently
terminated in mixed E-Sp mergers due to the presence of the fully grown super-massive BH 
in the E-progenitor, with lower mass ratio mergers (2:1) again resulting
in shorter termination timescales compared to higher mass ratio mixed (4:1)
mergers. Finally, dry E-E mergers with two super-massive BHs are extremely efficient in terminating any
residual star formation in the progenitors. Remergers of 1:1 merger remnants show typically no
star formation, whereas remergers of lower mass ratio merger remnants can show
star formation on the level of a few solar masses per year that is rapidly
terminated in the merger. This implies that already a second generation of dry merging is
expected to show very weak, if any, signatures of residual star formation.   

We show that in addition to 1:1 disk mergers \citep{2005Natur.433..604D} also 3:1 unequal-mass
disk mergers, mixed E-Sp mergers and dry E-E mergers reproduce the observed
$M_{\rm BH}-\sigma$ relation, with the dominant source of scatter arising from
variations in the initial gas mass fractions of the galaxies. The 3:1
unequal-mass mergers require BH repositioning, in particular at low numerical
resolution, which ensures that the final BH quickly resettles in the central
gas disk after the merging of the BHs. In addition, the  normalization
of the observed  $M_{\rm BH}-M_{*}$ relation is also well fitted in our
simulation sample, but with the resulting slope being slightly steeper compared to the
observed slope. This discrepancy might be resolved by employing a more aggressive
supernova-driven wind feedback, as in recent self-consistent cosmological simulations
\citep{2007MNRAS.380..877S,2008ApJ...676...33D} or by the fact the progenitors
of lower mass galaxy mergers have larger initial gas mass fractions compared
to high mass galaxy mergers. Finally, our
complete merger sample fits well the observed bulge binding energy correlation
with black hole mass, $M_{{\rm BH}}-M_{*}\sigma^{2}$, which is
statistically equivalent to the Black Hole Fundamental plane \citep{2007ApJ...669...45H,2007ApJ...669...67H}.
The tightness of this correlation indicates that the final BH mass in all the
merger samples is ultimately set by the depth of the gravitational potential well.

The remarkable robustness of the simulated $M_{\rm BH}-\sigma$, $M_{\rm
  BH}-M_{*}$ and $M_{{\rm BH}}-M_{*}\sigma^{2}$ relations for an extremely wide range of initial galaxy
properties implies that the \citet{2005MNRAS.361..776S}
effective BH feedback model self-regulates the mass growth of the
BHs in accordance with the observed relations for a very wide range of merger progenitors.
 However, in addition it is important to study the detailed properties of merger remnants
including BH feedback. It
has been shown that the stellar orbits, e.g. the relative population of box
and tube orbits, in merger remnants are affected by the amount of gas driven
to the center \citep{1996ApJ...471..115B,2005MNRAS.360.1185J,2007MNRAS.376..997J}.
Here, even relatively small amounts of gas ($\sim 5\%$) can influence
higher order photometric and kinematical properties of the remnants as a whole 
\citep{2006MNRAS.372..839N}.  As efficient feedback from black holes 
reduces the central gas content in first generation disk merger
remnants, future mergers will be less dissipative and favor the formation of
triaxial stellar systems dominated by box orbits
\citep{2005MNRAS.360.1185J,2006MNRAS.372..839N}. 
In addition, we studied in \citet{2007arXiv0710.0663B} the intrinsic anisotropy
$\delta$ and flattening $\epsilon$ of simulated merger remnants with 
and without BH feedback. Here, an equally good fit to ellipticals observed by SAURON
\citep{2004MNRAS.352..721E,2006MNRAS.366.1126C} was found in both models that included BH feedback and models
without BHs. However, binary disk mergers and their descendants were shown to
preferentially form anisotropic systems with relatively high ellipticities, 
not consistent with  observed isotropic and round ellipticals. This is kinematical 
evidence that some ellipticals cannot have formed by mergers of disks in the
first place. In addition alternative formation scenarios are motivated based
on stellar population considerations \citep{2007astro.ph..2535N} as well as
direct cosmological simulations \citep{2007ApJ...658..710N}. 

AGNs are emerging as an important ingredient in our current galaxy formation
models. Research on understanding the observed $M_{\rm BH}-\sigma$, $M_{\rm
  BH}-M_{*}$ and $M_{{\rm BH}}-M_{*}\sigma^{2}$ relations using numerical techniques should therefore
be a useful step in our quest to formulate a more complete theory of galaxy formation.

\begin{acknowledgements}

We would like to thank the anonymous referee for a careful reading of the
manuscript and valuable comments. In addition we thank V. Springel for helpful
comments and for supplying data for a comparison simulation. Furthermore, we thank 
P. Hopkins, R. Jesseit, J. Sommer-Larsen and M. Wetzstein for stimulating
discussions and helpful comments on the manuscript. The numerical simulations were performed on the local SGI-Altix
3700 Bx2, which was partly funded by the Cluster of Excellence: ''Origin and
Structure of the Universe''. 

\end{acknowledgements}

\nocite{*}
\bibliography{references}

\begin{thebibliography}{131}
\expandafter\ifx\csname natexlab\endcsname\relax\def\natexlab#1{#1}\fi

\bibitem[{{Abadi} {et~al.}(2003){Abadi}, {Navarro}, {Steinmetz}, \&
  {Eke}}]{2003ApJ...597...21A}
{Abadi}, M.~G., {Navarro}, J.~F., {Steinmetz}, M., \& {Eke}, V.~R. 2003, \apj,
  597, 21

\bibitem[{{Adams} {et~al.}(2001){Adams}, {Graff}, \&
  {Richstone}}]{2001ApJ...551L..31A}
{Adams}, F.~C., {Graff}, D.~S., \& {Richstone}, D.~O. 2001, \apjl, 551, L31

\bibitem[{{Aller} \& {Richstone}(2007)}]{2007ApJ...665..120A}
{Aller}, M.~C. \& {Richstone}, D.~O. 2007, \apj, 665, 120

\bibitem[{{Arad} \& {Johansson}(2005)}]{2005MNRAS.362..252A}
{Arad}, I. \& {Johansson}, P.~H. 2005, \mnras, 362, 252

\bibitem[{{Barger} {et~al.}(2005){Barger}, {Cowie}, {Mushotzky}, {Yang},
  {Wang}, {Steffen}, \& {Capak}}]{2005AJ....129..578B}
{Barger}, A.~J., {Cowie}, L.~L., {Mushotzky}, R.~F., {Yang}, Y., {Wang}, W.-H.,
  {Steffen}, A.~T., \& {Capak}, P. 2005, \aj, 129, 578

\bibitem[{{Barnes}(1998)}]{1998giis.conf..275B}
{Barnes}, J.~E. 1998, in Saas-Fee Advanced Course 26: Galaxies: Interactions
  and Induced Star Formation, ed. R.~C. {Kennicutt}, Jr., F.~{Schweizer}, J.~E.
  {Barnes}, D.~{Friedli}, L.~{Martinet}, \& D.~{Pfenniger}, 275--+

\bibitem[{{Barnes}(2002)}]{2002MNRAS.333..481B}
{Barnes}, J.~E. 2002, \mnras, 333, 481

\bibitem[{{Barnes} \& {Hernquist}(1996)}]{1996ApJ...471..115B}
{Barnes}, J.~E. \& {Hernquist}, L. 1996, \apj, 471, 115

\bibitem[{{Barway} \& {Kembhavi}(2007)}]{2007ApJ...662L..67B}
{Barway}, S. \& {Kembhavi}, A. 2007, \apjl, 662, L67

\bibitem[{{Bate} \& {Burkert}(1997)}]{1997MNRAS.288.1060B}
{Bate}, M.~R. \& {Burkert}, A. 1997, \mnras, 288, 1060

\bibitem[{{Bell} {et~al.}(2006){Bell}, {Naab}, {McIntosh}, {Somerville},
  {Caldwell}, {Barden}, {Wolf}, {Rix}, {Beckwith}, {Borch}, {H{\"a}ussler},
  {Heymans}, {Jahnke}, {Jogee}, {Koposov}, {Meisenheimer}, {Peng}, {Sanchez},
  \& {Wisotzki}}]{2006ApJ...640..241B}
{Bell}, E.~F., {Naab}, T., {McIntosh}, D.~H., {Somerville}, R.~S., {Caldwell},
  J.~A.~R., {Barden}, M., {Wolf}, C., {Rix}, H.-W., {Beckwith}, S.~V., {Borch},
  A., {H{\"a}ussler}, B., {Heymans}, C., {Jahnke}, K., {Jogee}, S., {Koposov},
  S., {Meisenheimer}, K., {Peng}, C.~Y., {Sanchez}, S.~F., \& {Wisotzki}, L.
  2006, \apj, 640, 241

\bibitem[{{Bondi}(1952)}]{1952MNRAS.112..195B}
{Bondi}, H. 1952, \mnras, 112, 195

\bibitem[{{Bondi} \& {Hoyle}(1944)}]{1944MNRAS.104..273B}
{Bondi}, H. \& {Hoyle}, F. 1944, \mnras, 104, 273

\bibitem[{{Bournaud} {et~al.}(2007{\natexlab{a}}){Bournaud}, {Elmegreen}, \&
  {Elmegreen}}]{2007ApJ...670..237B}
{Bournaud}, F., {Elmegreen}, B.~G., \& {Elmegreen}, D.~M. 2007{\natexlab{a}},
  \apj, 670, 237

\bibitem[{{Bournaud} {et~al.}(2005){Bournaud}, {Jog}, \&
  {Combes}}]{2005A&A...437...69B}
{Bournaud}, F., {Jog}, C.~J., \& {Combes}, F. 2005, \aap, 437, 69

\bibitem[{{Bournaud} {et~al.}(2007{\natexlab{b}}){Bournaud}, {Jog}, \&
  {Combes}}]{2007A&A...476.1179B}
---. 2007{\natexlab{b}}, \aap, 476, 1179

\bibitem[{{Bower} {et~al.}(2006){Bower}, {Benson}, {Malbon}, {Helly}, {Frenk},
  {Baugh}, {Cole}, \& {Lacey}}]{2006MNRAS.370..645B}
{Bower}, R.~G., {Benson}, A.~J., {Malbon}, R., {Helly}, J.~C., {Frenk}, C.~S.,
  {Baugh}, C.~M., {Cole}, S., \& {Lacey}, C.~G. 2006, \mnras, 370, 645

\bibitem[{{Boylan-Kolchin} {et~al.}(2005){Boylan-Kolchin}, {Ma}, \&
  {Quataert}}]{2005MNRAS.362..184B}
{Boylan-Kolchin}, M., {Ma}, C.-P., \& {Quataert}, E. 2005, \mnras, 362, 184

\bibitem[{{Bullock} {et~al.}(2001){Bullock}, {Kolatt}, {Sigad}, {Somerville},
  {Kravtsov}, {Klypin}, {Primack}, \& {Dekel}}]{2001MNRAS.321..559B}
{Bullock}, J.~S., {Kolatt}, T.~S., {Sigad}, Y., {Somerville}, R.~S.,
  {Kravtsov}, A.~V., {Klypin}, A.~A., {Primack}, J.~R., \& {Dekel}, A. 2001,
  \mnras, 321, 559

\bibitem[{{Burkert} \& {Naab}(2005)}]{2005MNRAS.363..597B}
{Burkert}, A. \& {Naab}, T. 2005, \mnras, 363, 597

\bibitem[{{Burkert} {et~al.}(2007){Burkert}, {Naab}, \&
  {Johansson}}]{2007arXiv0710.0663B}
{Burkert}, A., {Naab}, T., \& {Johansson}, P.~H. 2007, ArXiv e-prints, 710,
  arXiv:0710.0663

\bibitem[{{Burkert} \& {Silk}(2001)}]{2001ApJ...554L.151B}
{Burkert}, A. \& {Silk}, J. 2001, \apjl, 554, L151

\bibitem[{{Canalizo} \& {Stockton}(2001)}]{2001ApJ...555..719C}
{Canalizo}, G. \& {Stockton}, A. 2001, \apj, 555, 719

\bibitem[{{Cappellari} {et~al.}(2006){Cappellari}, {Bacon}, {Bureau}, {Damen},
  {Davies}, {de Zeeuw}, {Emsellem}, {Falc{\'o}n-Barroso}, {Krajnovi{\'c}},
  {Kuntschner}, {McDermid}, {Peletier}, {Sarzi}, {van den Bosch}, \& {van de
  Ven}}]{2006MNRAS.366.1126C}
{Cappellari}, M., {Bacon}, R., {Bureau}, M., {Damen}, M.~C., {Davies}, R.~L.,
  {de Zeeuw}, P.~T., {Emsellem}, E., {Falc{\'o}n-Barroso}, J., {Krajnovi{\'c}},
  D., {Kuntschner}, H., {McDermid}, R.~M., {Peletier}, R.~F., {Sarzi}, M., {van
  den Bosch}, R.~C.~E., \& {van de Ven}, G. 2006, \mnras, 366, 1126

\bibitem[{{Cattaneo} {et~al.}(1999){Cattaneo}, {Haehnelt}, \&
  {Rees}}]{1999MNRAS.308...77C}
{Cattaneo}, A., {Haehnelt}, M.~G., \& {Rees}, M.~J. 1999, \mnras, 308, 77

\bibitem[{{Ciotti} \& {Ostriker}(1997)}]{1997ApJ...487L.105C}
{Ciotti}, L. \& {Ostriker}, J.~P. 1997, \apjl, 487, L105+

\bibitem[{{Ciotti} \& {Ostriker}(2007)}]{2007ApJ...665.1038C}
---. 2007, \apj, 665, 1038

\bibitem[{{Ciotti} \& {van Albada}(2001)}]{2001ApJ...552L..13C}
{Ciotti}, L. \& {van Albada}, T.~S. 2001, \apjl, 552, L13

\bibitem[{{Cox} {et~al.}(2006){Cox}, {Dutta}, {Di Matteo}, {Hernquist},
  {Hopkins}, {Robertson}, \& {Springel}}]{2006ApJ...650..791C}
{Cox}, T.~J., {Dutta}, S.~N., {Di Matteo}, T., {Hernquist}, L., {Hopkins},
  P.~F., {Robertson}, B., \& {Springel}, V. 2006, \apj, 650, 791

\bibitem[{{Cox} {et~al.}(2008){Cox}, {Jonsson}, {Somerville}, {Primack}, \&
  {Dekel}}]{2008MNRAS.tmp...33C}
{Cox}, T.~J., {Jonsson}, P., {Somerville}, R.~S., {Primack}, J.~R., \& {Dekel},
  A. 2008, \mnras, 33

\bibitem[{{Cretton} {et~al.}(2001){Cretton}, {Naab}, {Rix}, \&
  {Burkert}}]{2001ApJ...554..291C}
{Cretton}, N., {Naab}, T., {Rix}, H.-W., \& {Burkert}, A. 2001, \apj, 554, 291

\bibitem[{{Croton} {et~al.}(2006){Croton}, {Springel}, {White}, {De Lucia},
  {Frenk}, {Gao}, {Jenkins}, {Kauffmann}, {Navarro}, \&
  {Yoshida}}]{2006MNRAS.365...11C}
{Croton}, D.~J., {Springel}, V., {White}, S.~D.~M., {De Lucia}, G., {Frenk},
  C.~S., {Gao}, L., {Jenkins}, A., {Kauffmann}, G., {Navarro}, J.~F., \&
  {Yoshida}, N. 2006, \mnras, 365, 11

\bibitem[{{Daddi} {et~al.}(2007{\natexlab{a}}){Daddi}, {Alexander},
  {Dickinson}, {Gilli}, {Renzini}, {Elbaz}, {Cimatti}, {Chary}, {Frayer},
  {Bauer}, {Brandt}, {Giavalisco}, {Grogin}, {Huynh}, {Kurk}, {Mignoli},
  {Morrison}, {Pope}, \& {Ravindranath}}]{2007ApJ...670..173D}
{Daddi}, E., {Alexander}, D.~M., {Dickinson}, M., {Gilli}, R., {Renzini}, A.,
  {Elbaz}, D., {Cimatti}, A., {Chary}, R., {Frayer}, D., {Bauer}, F.~E.,
  {Brandt}, W.~N., {Giavalisco}, M., {Grogin}, N.~A., {Huynh}, M., {Kurk}, J.,
  {Mignoli}, M., {Morrison}, G., {Pope}, A., \& {Ravindranath}, S.
  2007{\natexlab{a}}, \apj, 670, 173

\bibitem[{{Daddi} {et~al.}(2007{\natexlab{b}}){Daddi}, {Dickinson}, {Morrison},
  {Chary}, {Cimatti}, {Elbaz}, {Frayer}, {Renzini}, {Pope}, {Alexander},
  {Bauer}, {Giavalisco}, {Huynh}, {Kurk}, \& {Mignoli}}]{2007ApJ...670..156D}
{Daddi}, E., {Dickinson}, M., {Morrison}, G., {Chary}, R., {Cimatti}, A.,
  {Elbaz}, D., {Frayer}, D., {Renzini}, A., {Pope}, A., {Alexander}, D.~M.,
  {Bauer}, F.~E., {Giavalisco}, M., {Huynh}, M., {Kurk}, J., \& {Mignoli}, M.
  2007{\natexlab{b}}, \apj, 670, 156

\bibitem[{{Dasyra} {et~al.}(2006{\natexlab{a}}){Dasyra}, {Tacconi}, {Davies},
  {Genzel}, {Lutz}, {Naab}, {Burkert}, {Veilleux}, \&
  {Sanders}}]{2006ApJ...638..745D}
{Dasyra}, K.~M., {Tacconi}, L.~J., {Davies}, R.~I., {Genzel}, R., {Lutz}, D.,
  {Naab}, T., {Burkert}, A., {Veilleux}, S., \& {Sanders}, D.~B.
  2006{\natexlab{a}}, \apj, 638, 745

\bibitem[{{Dasyra} {et~al.}(2006{\natexlab{b}}){Dasyra}, {Tacconi}, {Davies},
  {Genzel}, {Lutz}, {Naab}, {Sanders}, {Veilleux}, \&
  {Baker}}]{2006NewAR..50..720D}
{Dasyra}, K.~M., {Tacconi}, L.~J., {Davies}, R.~I., {Genzel}, R., {Lutz}, D.,
  {Naab}, T., {Sanders}, D.~B., {Veilleux}, S., \& {Baker}, A.~J.
  2006{\natexlab{b}}, New Astronomy Review, 50, 720

\bibitem[{{Dasyra} {et~al.}(2006{\natexlab{c}}){Dasyra}, {Tacconi}, {Davies},
  {Naab}, {Genzel}, {Lutz}, {Sturm}, {Baker}, {Veilleux}, {Sanders}, \&
  {Burkert}}]{2006ApJ...651..835D}
{Dasyra}, K.~M., {Tacconi}, L.~J., {Davies}, R.~I., {Naab}, T., {Genzel}, R.,
  {Lutz}, D., {Sturm}, E., {Baker}, A.~J., {Veilleux}, S., {Sanders}, D.~B., \&
  {Burkert}, A. 2006{\natexlab{c}}, \apj, 651, 835

\bibitem[{{Dehnen}(2001)}]{2001MNRAS.324..273D}
{Dehnen}, W. 2001, \mnras, 324, 273

\bibitem[{{di Matteo} {et~al.}(2007){di Matteo}, {Combes}, {Melchior}, \&
  {Semelin}}]{2007A&A...468...61D}
{di Matteo}, P., {Combes}, F., {Melchior}, A.-L., \& {Semelin}, B. 2007, \aap,
  468, 61

\bibitem[{{Di Matteo} {et~al.}(2008){Di Matteo}, {Colberg}, {Springel},
  {Hernquist}, \& {Sijacki}}]{2008ApJ...676...33D}
{Di Matteo}, T., {Colberg}, J., {Springel}, V., {Hernquist}, L., \& {Sijacki},
  D. 2008, \apj, 676, 33

\bibitem[{{Di Matteo} {et~al.}(2005){Di Matteo}, {Springel}, \&
  {Hernquist}}]{2005Natur.433..604D}
{Di Matteo}, T., {Springel}, V., \& {Hernquist}, L. 2005, \nat, 433, 604

\bibitem[{{Efstathiou}(2000)}]{2000MNRAS.317..697E}
{Efstathiou}, G. 2000, \mnras, 317, 697

\bibitem[{{Elmegreen} {et~al.}(2008){Elmegreen}, {Bournaud}, \&
  {Elmegreen}}]{2008arXiv0805.2266E}
{Elmegreen}, B.~G., {Bournaud}, F., \& {Elmegreen}, D.~M. 2008, ArXiv e-prints,
  805, arXiv:0805.2266

\bibitem[{{Emsellem} {et~al.}(2004){Emsellem}, {Cappellari}, {Peletier},
  {McDermid}, {Bacon}, {Bureau}, {Copin}, {Davies}, {Krajnovi{\'c}},
  {Kuntschner}, {Miller}, \& {de Zeeuw}}]{2004MNRAS.352..721E}
{Emsellem}, E., {Cappellari}, M., {Peletier}, R.~F., {McDermid}, R.~M.,
  {Bacon}, R., {Bureau}, M., {Copin}, Y., {Davies}, R.~L., {Krajnovi{\'c}}, D.,
  {Kuntschner}, H., {Miller}, B.~W., \& {de Zeeuw}, P.~T. 2004, \mnras, 352,
  721

\bibitem[{{Escala} {et~al.}(2004){Escala}, {Larson}, {Coppi}, \&
  {Mardones}}]{2004ApJ...607..765E}
{Escala}, A., {Larson}, R.~B., {Coppi}, P.~S., \& {Mardones}, D. 2004, \apj,
  607, 765

\bibitem[{{Fabian}(1999)}]{1999MNRAS.308L..39F}
{Fabian}, A.~C. 1999, \mnras, 308, L39

\bibitem[{{Ferrarese} \& {Merritt}(2000)}]{2000ApJ...539L...9F}
{Ferrarese}, L. \& {Merritt}, D. 2000, \apjl, 539, L9

\bibitem[{{Flanagan} \& {Hughes}(1998)}]{1998PhRvD..57.4535F}
{Flanagan}, {\'E}.~{\'E}. \& {Hughes}, S.~A. 1998, \prd, 57, 4535

\bibitem[{{Gebhardt} {et~al.}(2000){Gebhardt}, {Bender}, {Bower}, {Dressler},
  {Faber}, {Filippenko}, {Green}, {Grillmair}, {Ho}, {Kormendy}, {Lauer},
  {Magorrian}, {Pinkney}, {Richstone}, \& {Tremaine}}]{2000ApJ...539L..13G}
{Gebhardt}, K., {Bender}, R., {Bower}, G., {Dressler}, A., {Faber}, S.~M.,
  {Filippenko}, A.~V., {Green}, R., {Grillmair}, C., {Ho}, L.~C., {Kormendy},
  J., {Lauer}, T.~R., {Magorrian}, J., {Pinkney}, J., {Richstone}, D., \&
  {Tremaine}, S. 2000, \apjl, 539, L13

\bibitem[{{Genzel} {et~al.}(1998){Genzel}, {Lutz}, {Sturm}, {Egami}, {Kunze},
  {Moorwood}, {Rigopoulou}, {Spoon}, {Sternberg}, {Tacconi-Garman}, {Tacconi},
  \& {Thatte}}]{1998ApJ...498..579G}
{Genzel}, R., {Lutz}, D., {Sturm}, E., {Egami}, E., {Kunze}, D., {Moorwood},
  A.~F.~M., {Rigopoulou}, D., {Spoon}, H.~W.~W., {Sternberg}, A.,
  {Tacconi-Garman}, L.~E., {Tacconi}, L., \& {Thatte}, N. 1998, \apj, 498, 579

\bibitem[{{Genzel} {et~al.}(2001){Genzel}, {Tacconi}, {Rigopoulou}, {Lutz}, \&
  {Tecza}}]{2001ApJ...563..527G}
{Genzel}, R., {Tacconi}, L.~J., {Rigopoulou}, D., {Lutz}, D., \& {Tecza}, M.
  2001, \apj, 563, 527

\bibitem[{{Granato} {et~al.}(2004){Granato}, {De Zotti}, {Silva}, {Bressan}, \&
  {Danese}}]{2004ApJ...600..580G}
{Granato}, G.~L., {De Zotti}, G., {Silva}, L., {Bressan}, A., \& {Danese}, L.
  2004, \apj, 600, 580

\bibitem[{{Haardt} \& {Madau}(1996)}]{1996ApJ...461...20H}
{Haardt}, F. \& {Madau}, P. 1996, \apj, 461, 20

\bibitem[{{H{\"a}ring} \& {Rix}(2004)}]{2004ApJ...604L..89H}
{H{\"a}ring}, N. \& {Rix}, H.-W. 2004, \apjl, 604, L89

\bibitem[{{Hernquist}(1989)}]{1989Natur.340..687H}
{Hernquist}, L. 1989, \nat, 340, 687

\bibitem[{{Hernquist}(1990)}]{1990ApJ...356..359H}
---. 1990, \apj, 356, 359

\bibitem[{{Hibbard} \& {Yun}(1999)}]{1999ApJ...522L..93H}
{Hibbard}, J.~E. \& {Yun}, M.~S. 1999, \apjl, 522, L93

\bibitem[{{Hopkins} {et~al.}(2008{\natexlab{a}}){Hopkins}, {Cox}, {Kere{\v s}},
  \& {Hernquist}}]{2008ApJS..175..390H}
{Hopkins}, P.~F., {Cox}, T.~J., {Kere{\v s}}, D., \& {Hernquist}, L.
  2008{\natexlab{a}}, \apjs, 175, 390

\bibitem[{{Hopkins} {et~al.}(2005){Hopkins}, {Hernquist}, {Cox}, {Di Matteo},
  {Martini}, {Robertson}, \& {Springel}}]{2005ApJ...630..705H}
{Hopkins}, P.~F., {Hernquist}, L., {Cox}, T.~J., {Di Matteo}, T., {Martini},
  P., {Robertson}, B., \& {Springel}, V. 2005, \apj, 630, 705

\bibitem[{{Hopkins} {et~al.}(2006){Hopkins}, {Hernquist}, {Cox}, {Di Matteo},
  {Robertson}, \& {Springel}}]{2006ApJS..163....1H}
{Hopkins}, P.~F., {Hernquist}, L., {Cox}, T.~J., {Di Matteo}, T., {Robertson},
  B., \& {Springel}, V. 2006, \apjs, 163, 1

\bibitem[{{Hopkins} {et~al.}(2008{\natexlab{b}}){Hopkins}, {Hernquist}, {Cox},
  \& {Kere{\v s}}}]{2008ApJS..175..356H}
{Hopkins}, P.~F., {Hernquist}, L., {Cox}, T.~J., \& {Kere{\v s}}, D.
  2008{\natexlab{b}}, \apjs, 175, 356

\bibitem[{{Hopkins} {et~al.}(2007{\natexlab{a}}){Hopkins}, {Hernquist}, {Cox},
  {Robertson}, \& {Krause}}]{2007ApJ...669...45H}
{Hopkins}, P.~F., {Hernquist}, L., {Cox}, T.~J., {Robertson}, B., \& {Krause},
  E. 2007{\natexlab{a}}, \apj, 669, 45

\bibitem[{{Hopkins} {et~al.}(2007{\natexlab{b}}){Hopkins}, {Hernquist}, {Cox},
  {Robertson}, \& {Krause}}]{2007ApJ...669...67H}
---. 2007{\natexlab{b}}, \apj, 669, 67

\bibitem[{{Houck} {et~al.}(2005){Houck}, {Soifer}, {Weedman}, {Higdon},
  {Higdon}, {Herter}, {Brown}, {Dey}, {Jannuzi}, {Le Floc'h}, {Rieke}, {Armus},
  {Charmandaris}, {Brandl}, \& {Teplitz}}]{2005ApJ...622L.105H}
{Houck}, J.~R., {Soifer}, B.~T., {Weedman}, D., {Higdon}, S.~J.~U., {Higdon},
  J.~L., {Herter}, T., {Brown}, M.~J.~I., {Dey}, A., {Jannuzi}, B.~T., {Le
  Floc'h}, E., {Rieke}, M., {Armus}, L., {Charmandaris}, V., {Brandl}, B.~R.,
  \& {Teplitz}, H.~I. 2005, \apjl, 622, L105

\bibitem[{{Hoyle} \& {Lyttleton}(1939)}]{1939PCPS...34..405H}
{Hoyle}, F. \& {Lyttleton}, R.~A. 1939, in Proceedings of the Cambridge
  Philisophical Society, Vol.~34, Proceedings of the Cambridge Philisophical
  Society, 405--+

\bibitem[{{Hughes}(2002)}]{2002MNRAS.331..805H}
{Hughes}, S.~A. 2002, \mnras, 331, 805

\bibitem[{{Jesseit} {et~al.}(2005){Jesseit}, {Naab}, \&
  {Burkert}}]{2005MNRAS.360.1185J}
{Jesseit}, R., {Naab}, T., \& {Burkert}, A. 2005, \mnras, 360, 1185

\bibitem[{{Jesseit} {et~al.}(2007){Jesseit}, {Naab}, {Peletier}, \&
  {Burkert}}]{2007MNRAS.376..997J}
{Jesseit}, R., {Naab}, T., {Peletier}, R.~F., \& {Burkert}, A. 2007, \mnras,
  376, 997

\bibitem[{{Johansson} \& {Efstathiou}(2006)}]{2006MNRAS.371.1519J}
{Johansson}, P.~H. \& {Efstathiou}, G. 2006, \mnras, 371, 1519

\bibitem[{{Johansson} {et~al.}(2004){Johansson}, {V{\"a}is{\"a}nen}, \&
  {Vaccari}}]{2004A&A...427..795J}
{Johansson}, P.~H., {V{\"a}is{\"a}nen}, P., \& {Vaccari}, M. 2004, \aap, 427,
  795

\bibitem[{{Katz} {et~al.}(1996){Katz}, {Weinberg}, \&
  {Hernquist}}]{1996ApJS..105...19K}
{Katz}, N., {Weinberg}, D.~H., \& {Hernquist}, L. 1996, \apjs, 105, 19

\bibitem[{{Kauffmann} \& {Haehnelt}(2000)}]{2000MNRAS.311..576K}
{Kauffmann}, G. \& {Haehnelt}, M. 2000, \mnras, 311, 576

\bibitem[{{Kazantzidis} {et~al.}(2005){Kazantzidis}, {Mayer}, {Colpi}, {Madau},
  {Debattista}, {Wadsley}, {Stadel}, {Quinn}, \& {Moore}}]{2005ApJ...623L..67K}
{Kazantzidis}, S., {Mayer}, L., {Colpi}, M., {Madau}, P., {Debattista}, V.~P.,
  {Wadsley}, J., {Stadel}, J., {Quinn}, T., \& {Moore}, B. 2005, \apjl, 623,
  L67

\bibitem[{{Kennicutt}(1998)}]{1998ARA&A..36..189K}
{Kennicutt}, Jr., R.~C. 1998, \araa, 36, 189

\bibitem[{{Khalatyan} {et~al.}(2008){Khalatyan}, {Cattaneo}, {Schramm},
  {Gottl{\"o}ber}, {Steinmetz}, \& {Wisotzki}}]{2008MNRAS.387...13K}
{Khalatyan}, A., {Cattaneo}, A., {Schramm}, M., {Gottl{\"o}ber}, S.,
  {Steinmetz}, M., \& {Wisotzki}, L. 2008, \mnras, 387, 13

\bibitem[{{Khochfar} \& {Burkert}(2001)}]{2001ApJ...561..517K}
{Khochfar}, S. \& {Burkert}, A. 2001, \apj, 561, 517

\bibitem[{{Khochfar} \& {Burkert}(2003)}]{2003ApJ...597L.117K}
---. 2003, \apjl, 597, L117

\bibitem[{{Khochfar} \& {Burkert}(2005)}]{2005MNRAS.359.1379K}
---. 2005, \mnras, 359, 1379

\bibitem[{{Khochfar} \& {Burkert}(2006)}]{2006A&A...445..403K}
---. 2006, \aap, 445, 403

\bibitem[{{Khochfar} \& {Silk}(2006)}]{2006MNRAS.370..902K}
{Khochfar}, S. \& {Silk}, J. 2006, \mnras, 370, 902

\bibitem[{{Kormendy} \& {Richstone}(1995)}]{1995ARA&A..33..581K}
{Kormendy}, J. \& {Richstone}, D. 1995, \araa, 33, 581

\bibitem[{{Levine} {et~al.}(2008){Levine}, {Gnedin}, {Hamilton}, \&
  {Kravtsov}}]{2008ApJ...678..154L}
{Levine}, R., {Gnedin}, N.~Y., {Hamilton}, A.~J.~S., \& {Kravtsov}, A.~V. 2008,
  \apj, 678, 154

\bibitem[{{Lynden-Bell}(1967)}]{1967MNRAS.136..101L}
{Lynden-Bell}, D. 1967, \mnras, 136, 101

\bibitem[{{Madau} \& {Quataert}(2004)}]{2004ApJ...606L..17M}
{Madau}, P. \& {Quataert}, E. 2004, \apjl, 606, L17

\bibitem[{{Magorrian} {et~al.}(1998){Magorrian}, {Tremaine}, {Richstone},
  {Bender}, {Bower}, {Dressler}, {Faber}, {Gebhardt}, {Green}, {Grillmair},
  {Kormendy}, \& {Lauer}}]{1998AJ....115.2285M}
{Magorrian}, J., {Tremaine}, S., {Richstone}, D., {Bender}, R., {Bower}, G.,
  {Dressler}, A., {Faber}, S.~M., {Gebhardt}, K., {Green}, R., {Grillmair}, C.,
  {Kormendy}, J., \& {Lauer}, T. 1998, \aj, 115, 2285

\bibitem[{{Makino} \& {Funato}(2004)}]{2004ApJ...602...93M}
{Makino}, J. \& {Funato}, Y. 2004, \apj, 602, 93

\bibitem[{{Marconi} \& {Hunt}(2003)}]{2003ApJ...589L..21M}
{Marconi}, A. \& {Hunt}, L.~K. 2003, \apjl, 589, L21

\bibitem[{{Mayer} {et~al.}(2007){Mayer}, {Kazantzidis}, {Madau}, {Colpi},
  {Quinn}, \& {Wadsley}}]{2007Sci...316.1874M}
{Mayer}, L., {Kazantzidis}, S., {Madau}, P., {Colpi}, M., {Quinn}, T., \&
  {Wadsley}, J. 2007, Science, 316, 1874

\bibitem[{{McKee} \& {Ostriker}(1977)}]{1977ApJ...218..148M}
{McKee}, C.~F. \& {Ostriker}, J.~P. 1977, \apj, 218, 148

\bibitem[{{McLure} {et~al.}(1999){McLure}, {Kukula}, {Dunlop}, {Baum}, {O'Dea},
  \& {Hughes}}]{1999MNRAS.308..377M}
{McLure}, R.~J., {Kukula}, M.~J., {Dunlop}, J.~S., {Baum}, S.~A., {O'Dea},
  C.~P., \& {Hughes}, D.~H. 1999, \mnras, 308, 377

\bibitem[{{Merritt} \& {Milosavljevi{\'c}}(2005)}]{2005LRR.....8....8M}
{Merritt}, D. \& {Milosavljevi{\'c}}, M. 2005, Living Reviews in Relativity, 8,
  8

\bibitem[{{Merritt} {et~al.}(2004){Merritt}, {Milosavljevi{\'c}}, {Favata},
  {Hughes}, \& {Holz}}]{2004ApJ...607L...9M}
{Merritt}, D., {Milosavljevi{\'c}}, M., {Favata}, M., {Hughes}, S.~A., \&
  {Holz}, D.~E. 2004, \apjl, 607, L9

\bibitem[{{Mo} {et~al.}(1998){Mo}, {Mao}, \& {White}}]{1998MNRAS.295..319M}
{Mo}, H.~J., {Mao}, S., \& {White}, S.~D.~M. 1998, \mnras, 295, 319

\bibitem[{{Monaghan}(1992)}]{1992ARA&A..30..543M}
{Monaghan}, J.~J. 1992, \araa, 30, 543

\bibitem[{{Murray} {et~al.}(2005){Murray}, {Quataert}, \&
  {Thompson}}]{2005ApJ...618..569M}
{Murray}, N., {Quataert}, E., \& {Thompson}, T.~A. 2005, \apj, 618, 569

\bibitem[{{Naab} \& {Burkert}(2003)}]{2003ApJ...597..893N}
{Naab}, T. \& {Burkert}, A. 2003, \apj, 597, 893

\bibitem[{{Naab} {et~al.}(1999){Naab}, {Burkert}, \&
  {Hernquist}}]{1999ApJ...523L.133N}
{Naab}, T., {Burkert}, A., \& {Hernquist}, L. 1999, \apjl, 523, L133

\bibitem[{{Naab} {et~al.}(2006{\natexlab{a}}){Naab}, {Jesseit}, \&
  {Burkert}}]{2006MNRAS.372..839N}
{Naab}, T., {Jesseit}, R., \& {Burkert}, A. 2006{\natexlab{a}}, \mnras, 372,
  839

\bibitem[{{Naab} {et~al.}(2007){Naab}, {Johansson}, {Ostriker}, \&
  {Efstathiou}}]{2007ApJ...658..710N}
{Naab}, T., {Johansson}, P.~H., {Ostriker}, J.~P., \& {Efstathiou}, G. 2007,
  \apj, 658, 710

\bibitem[{{Naab} {et~al.}(2006{\natexlab{b}}){Naab}, {Khochfar}, \&
  {Burkert}}]{2006ApJ...636L..81N}
{Naab}, T., {Khochfar}, S., \& {Burkert}, A. 2006{\natexlab{b}}, \apjl, 636,
  L81

\bibitem[{{Naab} \& {Ostriker}(2007)}]{2007astro.ph..2535N}
{Naab}, T. \& {Ostriker}, J.~P. 2007, ArXiv Astrophysics e-prints, 702, 0702535

\bibitem[{{Naab} \& {Trujillo}(2006)}]{2006MNRAS.369..625N}
{Naab}, T. \& {Trujillo}, I. 2006, \mnras, 369, 625

\bibitem[{{Navarro} {et~al.}(1997){Navarro}, {Frenk}, \&
  {White}}]{1997ApJ...490..493N}
{Navarro}, J.~F., {Frenk}, C.~S., \& {White}, S.~D.~M. 1997, \apj, 490, 493

\bibitem[{{Nipoti} {et~al.}(2003){Nipoti}, {Londrillo}, \&
  {Ciotti}}]{2003MNRAS.342..501N}
{Nipoti}, C., {Londrillo}, P., \& {Ciotti}, L. 2003, \mnras, 342, 501

\bibitem[{{Oppenheimer} \& {Dav{\'e}}(2006)}]{2006MNRAS.373.1265O}
{Oppenheimer}, B.~D. \& {Dav{\'e}}, R. 2006, \mnras, 373, 1265

\bibitem[{{Richstone} {et~al.}(1998){Richstone}, {Ajhar}, {Bender}, {Bower},
  {Dressler}, {Faber}, {Filippenko}, {Gebhardt}, {Green}, {Ho}, {Kormendy},
  {Lauer}, {Magorrian}, \& {Tremaine}}]{1998Natur.395A..14R}
{Richstone}, D., {Ajhar}, E.~A., {Bender}, R., {Bower}, G., {Dressler}, A.,
  {Faber}, S.~M., {Filippenko}, A.~V., {Gebhardt}, K., {Green}, R., {Ho},
  L.~C., {Kormendy}, J., {Lauer}, T.~R., {Magorrian}, J., \& {Tremaine}, S.
  1998, \nat, 395, A14+

\bibitem[{{Robertson} {et~al.}(2006{\natexlab{a}}){Robertson}, {Bullock},
  {Cox}, {Di Matteo}, {Hernquist}, {Springel}, \&
  {Yoshida}}]{2006ApJ...645..986R}
{Robertson}, B., {Bullock}, J.~S., {Cox}, T.~J., {Di Matteo}, T., {Hernquist},
  L., {Springel}, V., \& {Yoshida}, N. 2006{\natexlab{a}}, \apj, 645, 986

\bibitem[{{Robertson} {et~al.}(2006{\natexlab{b}}){Robertson}, {Cox},
  {Hernquist}, {Franx}, {Hopkins}, {Martini}, \&
  {Springel}}]{2006ApJ...641...21R}
{Robertson}, B., {Cox}, T.~J., {Hernquist}, L., {Franx}, M., {Hopkins}, P.~F.,
  {Martini}, P., \& {Springel}, V. 2006{\natexlab{b}}, \apj, 641, 21

\bibitem[{{Robertson} {et~al.}(2006{\natexlab{c}}){Robertson}, {Hernquist},
  {Cox}, {Di Matteo}, {Hopkins}, {Martini}, \&
  {Springel}}]{2006ApJ...641...90R}
{Robertson}, B., {Hernquist}, L., {Cox}, T.~J., {Di Matteo}, T., {Hopkins},
  P.~F., {Martini}, P., \& {Springel}, V. 2006{\natexlab{c}}, \apj, 641, 90

\bibitem[{{Rothberg} \& {Joseph}(2006)}]{2006AJ....132..976R}
{Rothberg}, B. \& {Joseph}, R.~D. 2006, \aj, 132, 976

\bibitem[{{Saslaw} {et~al.}(1974){Saslaw}, {Valtonen}, \&
  {Aarseth}}]{1974ApJ...190..253S}
{Saslaw}, W.~C., {Valtonen}, M.~J., \& {Aarseth}, S.~J. 1974, \apj, 190, 253

\bibitem[{{Shakura} \& {Syunyaev}(1973)}]{1973A&A....24..337S}
{Shakura}, N.~I. \& {Syunyaev}, R.~A. 1973, \aap, 24, 337

\bibitem[{{Sijacki} \& {Springel}(2006)}]{2006MNRAS.366..397S}
{Sijacki}, D. \& {Springel}, V. 2006, \mnras, 366, 397

\bibitem[{{Sijacki} {et~al.}(2007){Sijacki}, {Springel}, {di Matteo}, \&
  {Hernquist}}]{2007MNRAS.380..877S}
{Sijacki}, D., {Springel}, V., {di Matteo}, T., \& {Hernquist}, L. 2007,
  \mnras, 380, 877

\bibitem[{{Silk} \& {Rees}(1998)}]{1998A&A...331L...1S}
{Silk}, J. \& {Rees}, M.~J. 1998, \aap, 331, L1

\bibitem[{{Sillanp\"a\"a} {et~al.}(1988){Sillanp\"a\"a}, {Haarala}, {Valtonen},
  {Sundelius}, \& {Byrd}}]{1988ApJ...325..628S}
{Sillanp\"a\"a}, A., {Haarala}, S., {Valtonen}, M.~J., {Sundelius}, B., \&
  {Byrd}, G.~G. 1988, \apj, 325, 628

\bibitem[{{Springel}(2005)}]{2005MNRAS.364.1105S}
{Springel}, V. 2005, \mnras, 364, 1105

\bibitem[{{Springel} {et~al.}(2005{\natexlab{a}}){Springel}, {Di Matteo}, \&
  {Hernquist}}]{2005ApJ...620L..79S}
{Springel}, V., {Di Matteo}, T., \& {Hernquist}, L. 2005{\natexlab{a}}, \apjl,
  620, L79

\bibitem[{{Springel} {et~al.}(2005{\natexlab{b}}){Springel}, {Di Matteo}, \&
  {Hernquist}}]{2005MNRAS.361..776S}
---. 2005{\natexlab{b}}, \mnras, 361, 776

\bibitem[{{Springel} \& {Hernquist}(2002)}]{2002MNRAS.333..649S}
{Springel}, V. \& {Hernquist}, L. 2002, \mnras, 333, 649

\bibitem[{{Springel} \& {Hernquist}(2003)}]{2003MNRAS.339..289S}
---. 2003, \mnras, 339, 289

\bibitem[{{Springel} \& {Hernquist}(2005)}]{2005ApJ...622L...9S}
---. 2005, \apjl, 622, L9

\bibitem[{{Thomas} {et~al.}(2007){Thomas}, {Jesseit}, {Naab}, {Saglia},
  {Burkert}, \& {Bender}}]{2007MNRAS.381.1672T}
{Thomas}, J., {Jesseit}, R., {Naab}, T., {Saglia}, R.~P., {Burkert}, A., \&
  {Bender}, R. 2007, \mnras, 381, 1672

\bibitem[{{Tran} {et~al.}(2005){Tran}, {van Dokkum}, {Franx}, {Illingworth},
  {Kelson}, \& {Schreiber}}]{2005ApJ...627L..25T}
{Tran}, K.-V.~H., {van Dokkum}, P., {Franx}, M., {Illingworth}, G.~D.,
  {Kelson}, D.~D., \& {Schreiber}, N.~M.~F. 2005, \apjl, 627, L25

\bibitem[{{Tremaine} {et~al.}(2002){Tremaine}, {Gebhardt}, {Bender}, {Bower},
  {Dressler}, {Faber}, {Filippenko}, {Green}, {Grillmair}, {Ho}, {Kormendy},
  {Lauer}, {Magorrian}, {Pinkney}, \& {Richstone}}]{2002ApJ...574..740T}
{Tremaine}, S., {Gebhardt}, K., {Bender}, R., {Bower}, G., {Dressler}, A.,
  {Faber}, S.~M., {Filippenko}, A.~V., {Green}, R., {Grillmair}, C., {Ho},
  L.~C., {Kormendy}, J., {Lauer}, T.~R., {Magorrian}, J., {Pinkney}, J., \&
  {Richstone}, D. 2002, \apj, 574, 740

\bibitem[{{V{\"a}is{\"a}nen} {et~al.}(2008){V{\"a}is{\"a}nen}, {Mattila},
  {Kniazev}, {Adamo}, {Efstathiou}, {Farrah}, {Johansson}, {{\"O}stlin},
  {Buckley}, {Burgh}, {Crause}, {Hashimoto}, {Lira}, {Loaring}, {Nordsieck},
  {Romero-Colmenero}, {Ryder}, {Still}, \& {Zijlstra}}]{2008MNRAS.tmp..110V}
{V{\"a}is{\"a}nen}, P., {Mattila}, S., {Kniazev}, A., {Adamo}, A.,
  {Efstathiou}, A., {Farrah}, D., {Johansson}, P.~H., {{\"O}stlin}, G.,
  {Buckley}, D.~A.~H., {Burgh}, E.~B., {Crause}, L., {Hashimoto}, Y., {Lira},
  P., {Loaring}, N., {Nordsieck}, K., {Romero-Colmenero}, E., {Ryder}, S.,
  {Still}, M., \& {Zijlstra}, A. 2008, \mnras, 110

\bibitem[{{van Dokkum}(2005)}]{2005AJ....130.2647V}
{van Dokkum}, P.~G. 2005, \aj, 130, 2647

\bibitem[{{van Dokkum} {et~al.}(1999){van Dokkum}, {Franx}, {Fabricant},
  {Kelson}, \& {Illingworth}}]{1999ApJ...520L..95V}
{van Dokkum}, P.~G., {Franx}, M., {Fabricant}, D., {Kelson}, D.~D., \&
  {Illingworth}, G.~D. 1999, \apjl, 520, L95

\bibitem[{{Vitvitska} {et~al.}(2002){Vitvitska}, {Klypin}, {Kravtsov},
  {Wechsler}, {Primack}, \& {Bullock}}]{2002ApJ...581..799V}
{Vitvitska}, M., {Klypin}, A.~A., {Kravtsov}, A.~V., {Wechsler}, R.~H.,
  {Primack}, J.~R., \& {Bullock}, J.~S. 2002, \apj, 581, 799

\bibitem[{{Volonteri} {et~al.}(2003){Volonteri}, {Haardt}, \&
  {Madau}}]{2003ApJ...582..559V}
{Volonteri}, M., {Haardt}, F., \& {Madau}, P. 2003, \apj, 582, 559

\bibitem[{{Wyithe} \& {Loeb}(2003)}]{2003ApJ...595..614W}
{Wyithe}, J.~S.~B. \& {Loeb}, A. 2003, \apj, 595, 614

\end{thebibliography}

\appendix

\section{A. Black hole feedback model}
\label{BH_feedback}

The growth of the supermassive black holes is modeled following the
\citet{2005MNRAS.361..776S} model. In this effective
subresolution model the unresolved accretion onto the BH is related
to the resolved gas distribution using a Bondi-Hoyle-Lyttleton parameterization
\citep{1939PCPS...34..405H,1944MNRAS.104..273B,1952MNRAS.112..195B}. In this description, the accretion rate onto the BH is given by

\begin{equation}
\dot{M}_{\rm{B}}=\frac{4 \pi \alpha G^{2} M_{\rm BH}^{2} \rho}{(c_{\rm s}^2+v^{2})^{3/2}}.
\label{Bondi}
\end{equation}
Here $\rho$ and $c_{\rm s}$ are the density and sound speed of the
surrounding gas, respectively, $v$ is the velocity of the BH relative to the
surrounding gas and $\alpha$ is a dimensionless parameter setting the efficiency of the
accretion. In addition, it is assumed that the maximum accretion is limited to
the Eddington rate

\begin{equation}
\dot{M}_{\rm{edd}}=\frac{4 \pi G M_{\rm BH} m_{\rm{p}}}{\epsilon_{\rm{r}}\sigma_{\rm{T}} c},
\label{Eddington}
\end{equation}
where $m_{\rm{p}}$ is the proton mass,  $\sigma_{\rm{T}}$ is the Thomson
cross-section and $\epsilon_{\rm{r}}$ is the radiative efficiency, that we
assume to be 0.1 based on the mean value for 
radiatively efficient \cite{1973A&A....24..337S} accretion onto a Schwarzschild BH.
Thus the accretion rate is given by
$\dot{M}_{\rm{BH}}=\rm{min}(\dot{M}_{\rm{B}},\dot{M}_{\rm{edd}})$.
Finally some fraction $\epsilon_{\rm{f}}$ of the radiated
luminosity $L_{\rm{r}}$ couples thermally and isotropically to the surrounding
gas in the form of black hole feedback energy resulting in

\begin{equation}
\dot{E}_{\rm feed}=\epsilon_{f} L_{\rm{r}}= \epsilon_{f} \epsilon_{\rm{r}} \dot{M}_{\rm{BH}}c^{2}.
\end{equation}
Again, following \citet{2005MNRAS.361..776S}, we fix $\epsilon_{f}=0.05$ so that
in total 0.5 per cent of the accreted rest mass energy is available as
feedback. This choice fixes the normalization of the $M_{\rm BH}-\sigma$ 
relation and reproduces the observed relation in equal-mass binary
galaxy merger simulations with BH feedback as was shown by \citet{2005Natur.433..604D}. The feedback energy is 
distributed as thermal energy to the surrounding gas particles weighted with
the SPH kernel. Numerically the BHs are represented as collisionless 'sink' particles, which
only feel gravitational forces. The properties, such as density, temperature
and the bulk velocity of the local gas around the BH are estimated in a similar 
fashion to normal SPH particles. Based on these quantities and the equations
above, the BH accretion rate is then estimated.

\section{B. Comparison of the BH feedback models}
\label{volker_comp}

\begin{figure}
\centering 
\includegraphics[width=9cm]{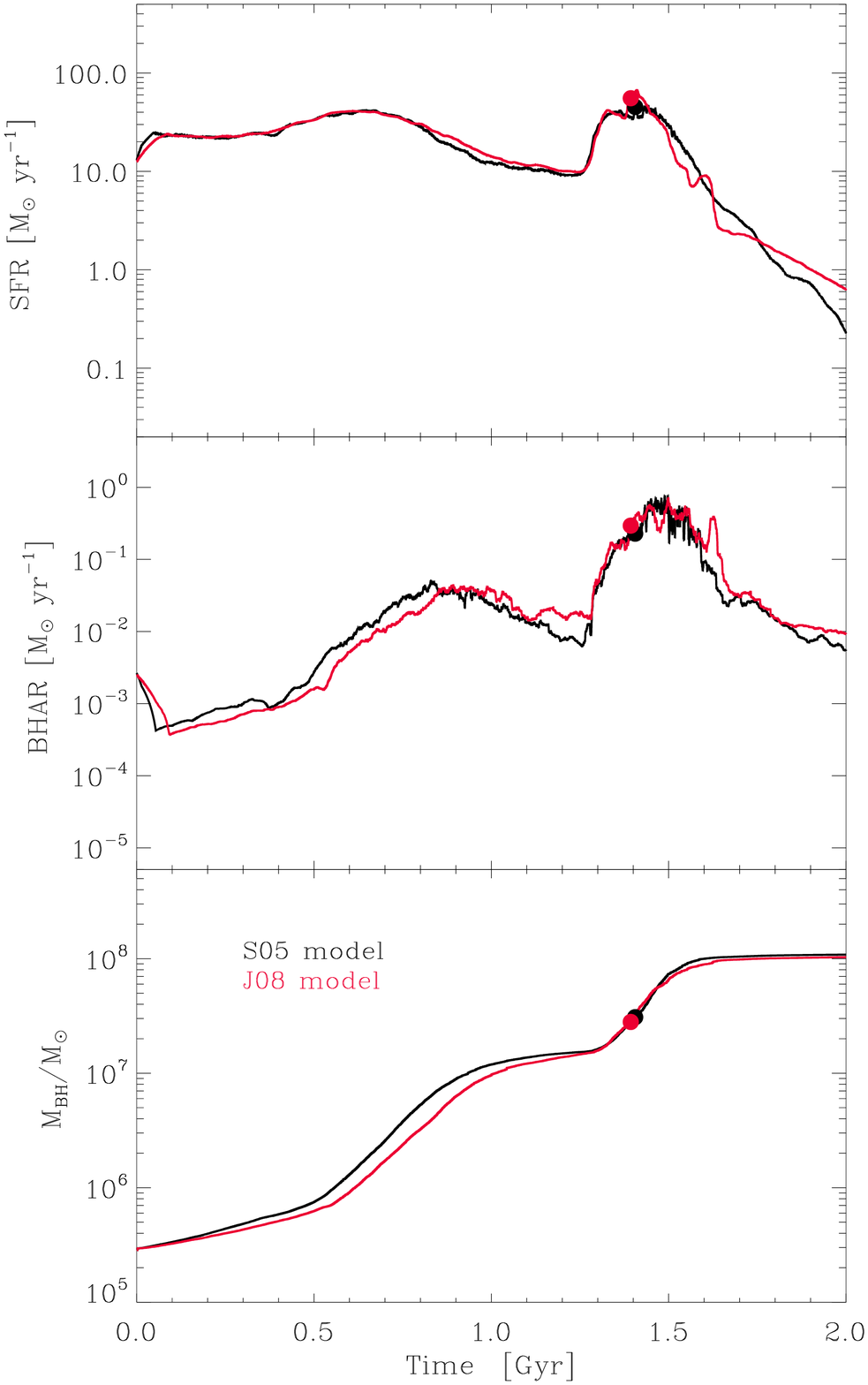}
\caption{Comparison of the total star formation rate (top), the total black hole accretion rate (middle)
and the evolution of the total black hole mass (bottom) as a function of time
for a 1:1 merger run using the \citet{2005MNRAS.361..776S} BH feedback
S05 model and the model presented in this paper (J08 model). The filled
circles indicate the time of merging of the BHs.}
\label{Code_comp}
\end{figure}

We perform a direct comparison of our BH feedback model to the \citet{2005MNRAS.361..776S}
model by running one of the initial conditions used in the
\citet{2005MNRAS.361..776S} study (kindly provided by Volker Springel) using
our simulation code. We use identical numerical parameters governing the
stellar feedback, BH feedback and integration accuracy. In
Fig. \ref{Code_comp} we plot the resulting star formation rate, BH accretion
rate and total BH mass for the simulation performed using the
\citet{2005MNRAS.361..776S} (S05 model, data provided by Volker Springel) and
compare it to the corresponding output of the model presented in this paper (J08 model).

The agreement between the two codes is very good, especially given the fact
that both the star formation and BH accretion processes are treated
stochastically. The star formation histories agree very well 
up to the time of the merging of the BHs. After the merging of the BHs we see small
differences in the decay rate of star formation in the two models.  
This slight discrepancy is due to the differences in the corresponding BH
accretion rates resulting in the final BH being marginally more massive in the
S05 model. The late decay of star formation is relatively sensitive to the
final BH mass, with more massive BHs resulting in more
efficient termination of star formation and hence small differences in the
late star formation history are not unexpected.

We thus conclude given the stochastic nature of the
modeled processes that the resulting agreement on the 10\% level between the two codes 
is very good. Thus we expect our BH feedback prescription to produce nearly
identical results compared to the original \citet{2005MNRAS.361..776S} model.

\section{C. Comparison of the merger prescriptions}
\label{merg_pre_res}

\begin{figure*}
\centering 
\includegraphics[width=20cm]{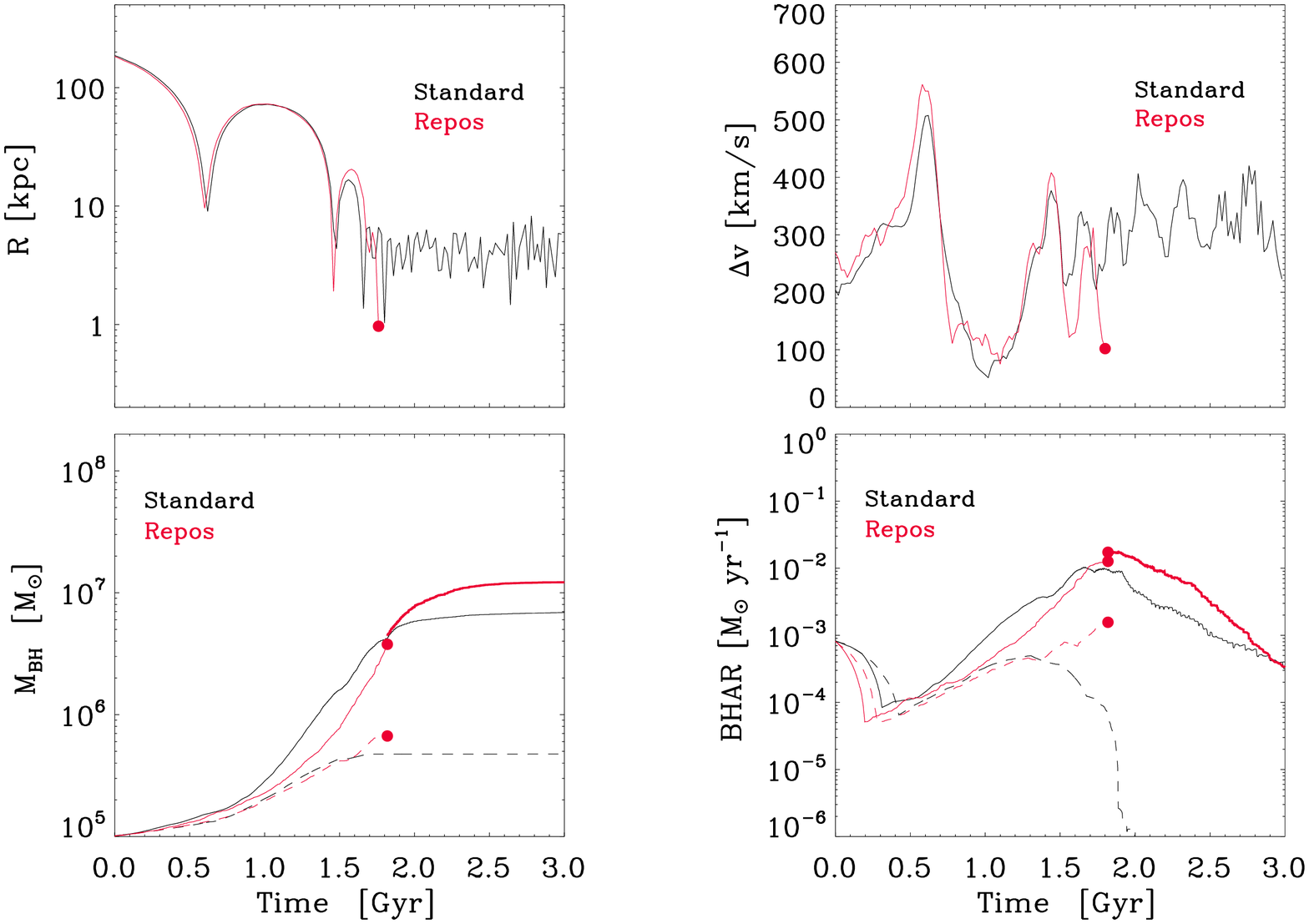}
\caption{3:1 unequal-mass merger simulations run at the standard numerical
  resolution with the standard BH merger prescription (black lines) and the
  repositioning method (red lines). The top left and right panels shows the relative
  distances and velocities between the BHs as a function of time. The bottom panels show the
  evolution of the BH masses and accretion rates as a function of time. The
  solid line depicts the primary BH, the dashed line the secondary BH and the
  thick solid line the merged BH, with the time of merging indicated with 
  solid circles.}
\label{31_details}
\end{figure*}

\begin{figure*}
\centering 
\includegraphics[width=20cm]{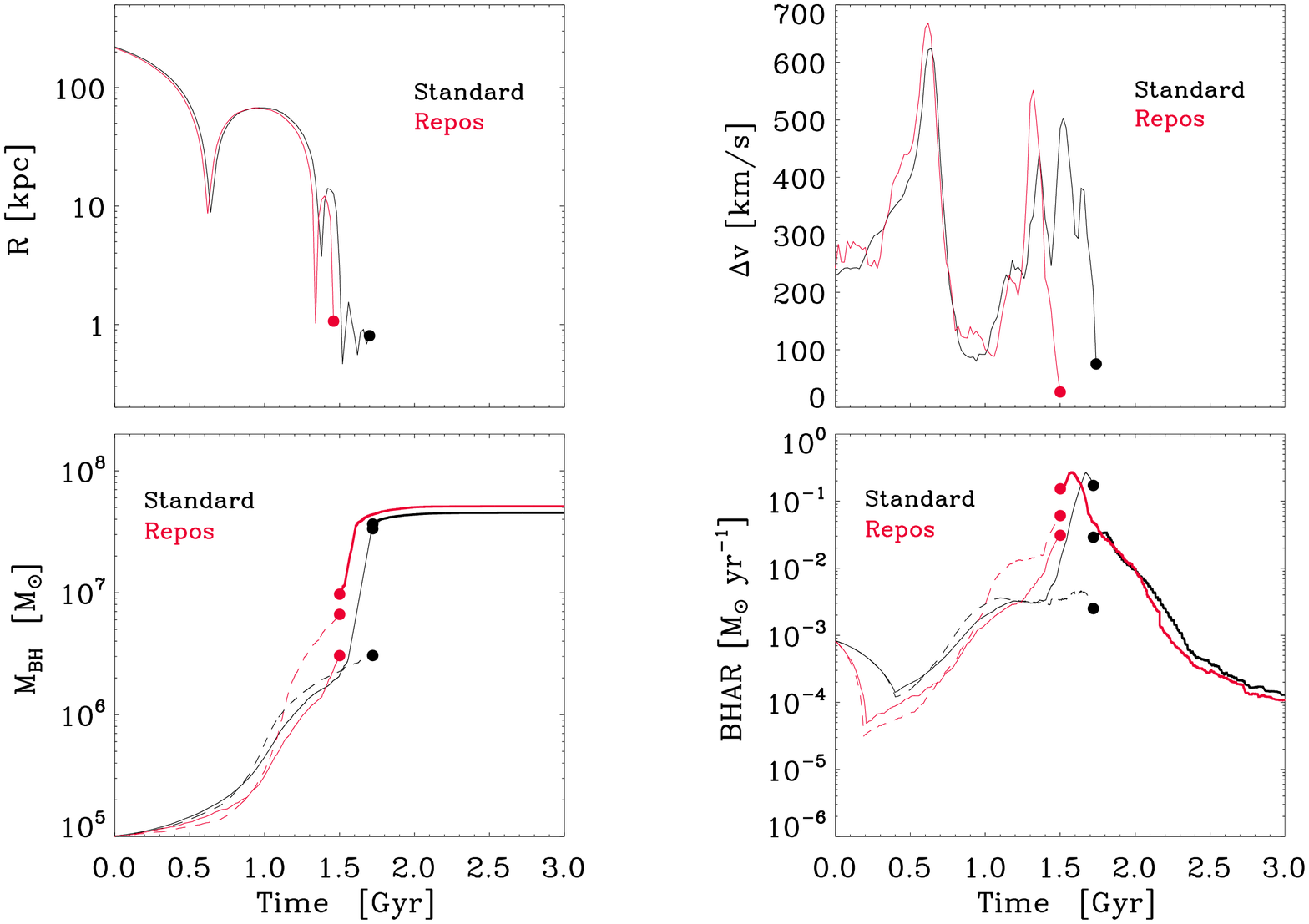}
\caption{1:1 equal-mass merger simulations run at the standard numerical
  resolution with the standard BH merger prescription (black lines) and the
  repositioning method (red lines). The top left and right panels shows the relative
  distances and velocities between the BHs as a function of time. The bottom panels show the
  evolution of the BH masses and accretion rates as a function of time. The
  solid line depicts the primary BH, the dashed line the secondary BH and the
  thick solid line the merged BH, with the time of merging indicated with 
  solid circles.}
\label{11_details}
\end{figure*}

We plot in Fig. \ref{31_details} the black hole mass growth and
mass accretion rate together with the distance and relative velocity between
the black holes for one unequal-mass 3:1 merger at our standard resolution using both 
the standard and repositioning merger prescriptions.
The galaxies go through their first and second passages at
$t_{1}=0.6 \ \rm{Gyr}$ and $t_{2}=1.4 \ \rm{Gyr}$ respectively. The galaxies then
approach each other for a third time at $t_{3}=1.8 \ \rm{Gyr}$ at which time the
BHs in the simulation with BH repositioning merge, unlike the standard case in
which the BHs do not merge. The mass growth of both the primary (solid lines)
and the secondary (dashed lines) BHs are similar for both methods until the
merger, after which time essentially only the repositioned BH grows in mass. 
A similar result can be seen in the BH accretion rates, which are similar
until  $t_{3}=1.8 \ \rm{Gyr}$, after which the repositioned BH shows almost an
order of magnitude higher accretion rates compared to the standard BH merger
case. One can also see that the secondary BH mass accretion rate dramatically
drops at this time, hence terminating the growth of 
the secondary BH. This is caused by the fact that
the secondary BH is perturbed away from its surrounding gas disk during 
the final stages of the merger. 

In the standard BH merger prescription the merging of the BHs is typically 
delayed with respect to the BHs in the repositioning runs. During the BH merger the 
momentum is conserved, 
which can give the BH a relatively large 'kick', especially in the 3:1 mergers
where the mass ratio of the BHs is relatively large. The repositioning ensures
that after the merger the BH quickly resettles in the center of the central
gas disk that is feeding the BH. In the standard prescription the merged BH
remains longer displaced from the central gas disk, thus lowering the mass
accretion rate, which scales linearly with density ($\dot{M}_{\rm{B}}\propto
\rho$ in Eq. \ref{Bondi}). The repositioning also lowers the relative
velocity of the BH with respect to the surrounding gas. For the
standard merging case the relative velocity can be relatively large with the BH plunging through
the central gaseous disk. This also affects the
BH accretion rate as it scales as the inverse third power of the relative
BH-gas velocity for a fixed sound speed of the medium ($\dot{M}_{\rm{B}}\propto
v^{-3}$ in Eq. \ref{Bondi}). Finally the BH accretion rate scales with the
square of the BH mass ($\dot{M}_{\rm{B}}\propto
M_{\rm BH}^{2}$ in Eq. \ref{Bondi}) resulting in higher accretion rates for
more massive BHs, thus an earlier merging of the BHs produces typically larger
final BH masses.

In Fig. \ref{11_details} we plot the corresponding details for a 1:1 merger at
our standard numerical resolution using the two merging prescriptions. The
differences between the merger prescriptions
compared to the 3:1 merger are less pronounced with the BHs merging slightly
earlier in the repositioned simulation. However, the growth of the black hole
masses show differences with the BHs in the repositioned run showing roughly
equal mass growth prior to the BH merging compared to the standard case, where 
the more massive BH is an
order of magnitude more massive than the secondary BH at the time of the
BH merger. In the standard merging case the secondary BH is perturbed away from
its gaseous central disk prior to the final merger, thus lowering the
accretion rate and growth of the BH in comparison to the primary BH. After the
merger the repositioned BH exhibits strong growth in a gas-rich environment
until the thermal feedback energy input is large enough to clear away the remaining gas,
thus terminating the BH growth. This is not mirrored in the standard merging
case, where the primary BH grows strongly prior to merging and only modestly
after the merger due to the fact that the merged BH is unable to resettle in
the remaining central gas disk. However, the end result of both merging
prescriptions is a final BH mass and stellar velocity dispersion that agree
with each other on the 10\% level.   

Taken together these results indicate that repositioning the BHs produce converged
and stable results for the mass growth and accretion rates of the
BHs. Repositioning is essential in representing the evolution of BHs in 3:1 mergers
and also produces converged results for 1:1 mergers even at relatively low
numerical resolution.

\end{document}